\documentclass[11pt,envcountsame,envcountsect,runningheads]{llncs}
\usepackage[a4paper,margin=2.5cm]{geometry}

\newcommand{\ShortVersion}[1]{} 
\newcommand{\LongVersion}[1]{#1} 
\newcommand{\VSpace}[1]{} 

\usepackage{latexsym} 
\usepackage{xspace} 
\usepackage{amssymb} 
\usepackage{amsmath}
\usepackage{stmaryrd} 
\usepackage{url} 
\usepackage{oldgerm} 
\input{xy} 
\xyoption{all} 
\usepackage[ruled,vlined,linesnumbered]{algorithm2e}

\newcommand{\V}{\forall}
\newcommand{\E}{\exists}
\newcommand{\Dmd}{\Diamond}
\newcommand{\lDmd}[1]{\langle#1\rangle}

\newcommand{\ovl}[1]{\overline{#1}}
\newcommand{\fracc}[2]{\displaystyle{\frac{\;#1\;}{\;#2\;}}}

\newcommand{\EXPTIME}{\textsc{Exp\-Time}\xspace}

\newcommand{\mL}{\mathcal{L}}

\def\trojkat{\mbox{{\scriptsize$\!\vartriangleleft$}}}

\def\defeq{=}

\def\eqref#1{(\ref{#1})}

\def\trojkat{\mbox{{\scriptsize$\!\vartriangleleft$}}}
\newcommand{\myEnd}{\mbox{}\hfill\trojkat}

\let\oldmarginpar\marginpar
\renewcommand\marginpar[1]{\oldmarginpar{\center{\footnotesize{\em #1}}}}

\newcommand{\comment}[1]{} 

\newenvironment{Definition}{\begin{definition}\em}{\myEnd\end{definition}}

\newcommand{\CPDLreg}{CPDL$_{reg}$\xspace}
\newcommand{\CPDL}{CPDL\xspace}
\newcommand{\CReg}{REG$^c$\xspace}

\newcommand{\SROIQ}{$\mathcal{SROIQ}$\xspace}
\newcommand{\CIQ}{$\mathcal{CIQ}$\xspace}
\newcommand{\ALC}{$\mathcal{ALC}$\xspace}

\newcommand{\rAnd}{(\land)}
\newcommand{\rOr}{(\lor)}
\newcommand{\rAutB}{(aut_\Box)}
\newcommand{\rAutD}{(aut_\Dmd)}
\newcommand{\rBox}{([\mathrm{A}])}
\newcommand{\rBoxD}{([\mathrm{A}]_f)}
\newcommand{\rDmd}{(\lDmd{\mathrm{A}})}
\newcommand{\rDmdD}{(\lDmd{\mathrm{A}}_f)}
\newcommand{\rBoxQm}{(\Box_{?})}
\newcommand{\rDmdQm}{(\Dmd_{?})}
\newcommand{\rTrans}{(trans)}

\newcommand{\rConv}{(conv)}
\newcommand{\rFormingState}{(forming\textrm{-}state)}

\newcommand{\cnv}[1]{{#1}^-}
\newcommand{\props}{\mathcal{PROP}}
\newcommand{\mindices}{\Sigma}

\newcommand{\clsL}{\mathit{closure}_L}
\newcommand{\bsf}{\mathit{bsf}}

\newcommand{\aut}{\mathbf{A}}
\newcommand{\Aut}{\mathbb{A}}

\newcommand{\cL}{$\mathcal{C}\!\mathit{L}$\xspace}

\newcommand{\Type}{\mathit{Type}}
\newcommand{\Status}{\mathit{Status}}
\newcommand{\Label}{\mathit{Label}}
\newcommand{\CELabel}{\mathit{CELabel}}
\newcommand{\StatePred}{\mathit{StatePred}}
\newcommand{\AfterTransPred}{\mathit{ATPred}}
\newcommand{\RFormulas}{\mathit{RFmls}}
\newcommand{\DFormulas}{\mathit{DFmls}}
\newcommand{\DSTimeStamp}{\mathit{DSTimeStamp}}
\newcommand{\FormulaSC}{\mathit{FmlSC}}

\newcommand{\SemiEndNodes}{\mathit{SemiEndNodes}}

\newcommand{\State}{\mathsf{state}}
\newcommand{\NonState}{\mathsf{non\textrm{-}state}}
\newcommand{\AndNode}{\mathsf{and\textrm{-}node}}
\newcommand{\OrNode}{\mathsf{or\textrm{-}node}}
\newcommand{\True}{\mathsf{true}}
\newcommand{\False}{\mathsf{false}}
\newcommand{\Sat}{\mathsf{sat}}
\newcommand{\Unsat}{\mathsf{unsat}}
\newcommand{\Incomplete}{\mathsf{incomplete}}
\newcommand{\Expanded}{\mathsf{expanded}}
\newcommand{\Unexpanded}{\mathsf{unexpanded}}
\newcommand{\Null}{\mathsf{null}}

\SetKwInput{Input}{Input}
\SetKwInput{Output}{Output}
\SetKwInput{GlobalData}{Global data}
\SetKwInput{Purpose}{Purpose}
\SetKw{Call}{call}
\SetKw{Set}{set}
\SetKw{Exit}{exit}
\SetKw{Break}{break}
\SetKw{Continue}{continue}

\SetKwFunction{NewSucc}{NewSucc} 
\SetKwFunction{FindProxy}{FindProxy} 
\SetKwFunction{ConToSucc}{ConToSucc} 
\SetKwFunction{Trans}{Trans}
\SetKwFunction{Apply}{Apply}
\SetKwFunction{Tableau}{Tableau}
\SetKwFunction{Expand}{Expand}
\SetKwFunction{UpdateStatus}{UpdateStatus}
\SetKwFunction{UpdateUnsatNodes}{UpdateUnsatNodes}
\SetKwFunction{PropagateStatus}{PropagateStatus}
\SetKwFunction{Kind}{Kind}
\SetKwFunction{AfterTrans}{AfterTrans}
\SetKwFunction{FullLabel}{FullLabel}
\SetKwFunction{BeforeFormingState}{BeforeFormingState}
\SetKwFunction{TUnsat}{TUnsat}
\SetKwFunction{TSat}{TSat}
\SetKwFunction{ToExpand}{ToExpand}
\SetKwFunction{AFormulas}{AFmls}
\SetKwFunction{NDFormulas}{NDFmls}

\title{A Cut-Free \EXPTIME Tableau Decision Procedure for the Logic Extending Converse-PDL with Regular Inclusion Axioms\thanks{This work is supported by Polish National Science Centre grant 2011/01/B/ST6/02759.}} 

\titlerunning{Cut-Free \EXPTIME Tableaux for \CPDLreg} 

\author{Linh Anh Nguyen} 
\institute{ 
Institute of Informatics, University of Warsaw\\ 
Banacha 2, 02-097 Warsaw, Poland\\ 
\email{nguyen@mimuw.edu.pl} 
} 

\authorrunning{L.A. Nguyen}	 

\tocauthor{Linh Anh Nguyen (University of Warsaw)} 

\begin{document} 
\maketitle  
\sloppy 

\begin{abstract}
We give the first cut-free \EXPTIME (optimal) tableau decision procedure for the logic \CPDLreg, which extends Converse-PDL with regular inclusion axioms characterized by finite automata. The logic \CPDLreg is the combination of Converse-PDL and regular grammar logic with converse. Our tableau decision procedure uses global state caching and has been designed to increase efficiency and allow various optimization techniques, including on-the-fly propagation of local and global (in)consistency. 
\end{abstract}

\section{Introduction}

In this paper we study automated reasoning in the modal logic \CPDLreg, which is a combination of \CPDL (propositional dynamic logic with converse \cite{HKT00,de-giacomo-massacci-converse-pdl,NguyenSzalas-KSE09}) and \CReg (regular grammar logic with converse \cite{ddn-jlli05,NguyenSzalas-CADE-22}). The logic \CPDL is widely used in many areas, including program verification, theory of action and change as well as knowledge representation (see, e.g., \cite{HKT00,GiacomoL96,HSPDL}). The logic \CReg can be used for modeling and reasoning about epistemic states of multi-agent systems~\cite{GoreNguyenCLIMA07} and Web ontologies~\cite{NguyenSzalas-CADE-22,HorrocksKS06}. 

In the papers~\cite{GBGI,dkns2011} written jointly with Dunin-K\c{e}plicz and Sza{\l}as, we show that \CPDLreg is a logical formalism suitable for expressing complex properties of agents' cooperation in terms of beliefs, goals and intentions. 
The logic \CPDLreg can also be used as a description logic (DL). As a combination of CPDL and \CReg, it allows to use role constructors (by program constructors of \CPDL) like the DL \CIQ~\cite{GiacomoL96} and to use role inclusion axioms (by inclusion axioms of \CReg) similarly to the DL \SROIQ~\cite{HorrocksKS06} (but by using automata instead of grammar rules). 

The mentioned works~\cite{GBGI,dkns2011} present a tableau calculus leading to the first \EXPTIME (optimal) tableau decision procedure for \CPDLreg. 
Observing the rules of that calculus, it can be seen that the general satisfiability checking problem in \CPDLreg is reducible to the satisfiability checking problem in \CPDL. 
However, translating the finite automata of a given \CPDLreg logic $L$ into regular expressions and then applying an \EXPTIME (optimal) decision procedure like the ones given by us and Sza{\l}as~\cite{NguyenSzalas-KSE09} or by Gor{\'e} and Widmann~\cite{GoreW10} for the resulted satisfiability problem in \CPDL may require double exponential time in the size of the original problem. The reason is that the regular expressions resulted from translating the finite automata specifying $L$ may have exponential lengths. 
One can also consider the method of translating \CReg into \CPDL given in~\cite{DemriNivelle2004} by Demri and de Nivelle. However, as \CPDLreg is much more complicated than \CReg, it is not trivial at all how to generate that method for translating \CPDLreg into \CPDL and how efficient the approach would be. 
Therefore, as stated in~\cite{dkns2011}, it is worth studying the direct approach for automated reasoning in \CPDLreg. 

The tableau calculus given in our joint papers~\cite{GBGI,dkns2011} for \CPDLreg uses analytic cuts and therefore is not efficient in practice. 
In this paper we improve that calculus by eliminating cuts to give the first cut-free \EXPTIME (optimal) tableau decision procedure for \CPDLreg. Our calculus uses global state caching as in the works~\cite{GoreW09,GoreW10} by Gor{\'e} and Widmann. It also uses local caching for non-states of tableaux. 

The idea of global caching comes from Pratt's paper on PDL~\cite{Pratt80}. In~\cite{GoreNguyen05tab,GoreNguyenCLIMA07,GoreNguyenTab07} together with Gor{\'e} we formalized and applied it to the modal logics REG (regular grammar logics), BReg (regular modal logics of agent beliefs) and the description logic $\mathcal{SHI}$ to obtain \EXPTIME (optimal) tableau decision procedures for these logics. Later, together with Sza\l{}as we extended the method to give \EXPTIME (optimal) tableau decision procedures for the logics PDL~\cite{NguyenS10FI}, CPDL~\cite{NguyenSzalas-KSE09} and \CReg~\cite{NguyenSzalas-CADE-22}. In~\cite{GoreNguyenTab07,NguyenSzalas-KSE09,NguyenSzalas-CADE-22} we used analytic cuts to deal with inverse roles and converse modal operators. As cuts are not efficient in practice, Gor{\'e} and Widmann developed the first cut-free \EXPTIME tableau decision procedures, based on global state caching, for the DL $\mathcal{ALCI}$~\cite{GoreW09} and CPDL~\cite{GoreW10}. 

In the current paper, we use the idea of global state caching and a slightly different technique to deal with converse modal operators for \CPDLreg. We check ``compatibility'' of a non-state $w$ with its predecessor-state $v$ as soon as possible and do not require $w$ to be ``saturated''. In~\cite{GoreW09,GoreW10} Gor{\'e} and Widmann require $w$ to be ``saturated'' before checking the compatibility, which has a serious drawback, for example, when the label of $v$ is of the form $\{p$, $\lDmd{\sigma}[\sigma^-]\lnot p$, $[\sigma]((p_1 \lor q_1) \land\ldots\land(p_n \lor q_n))\}$. Besides, we tend to give a higher priority for the current branch even when it involves converse, while Gor{\'e} and Widmann~\cite{GoreW09,GoreW10} tend to delay solving incompatibility w.r.t. converse.\footnote{We use the verb ``tend'' because any systematic search strategy is applicable for both \cite{GoreW09,GoreW10} and the current paper. Nevertheless, the ``tendency'' can still be observed.} Our technique of dealing with converse modal operators for \CPDLreg can be described as follows: if a state $v$ is ``\textsf{toosmall}'' due to the lack of a formula $\varphi$ and a non-state $u$ is the unique predecessor of $v$ then delete the edge $(u,v)$ and connect $u$ to two nodes $v_1$ and $v_2$ (which are created when necessary) such that $v_1$ extends $v$ with $\varphi$ and $v_2$ is similar to $v$ but always ``disallows'' $\varphi$ (i.e. if any non-state $w$ having $v_2$ as the predecessor-state requires adding $\varphi$ to $v_2$ then we give $w$ status $\Unsat$). For simplicity, in the current version of this paper we do not consider separately the case when $v$ is ``\textsf{toosmall}'' due to the lack of a set $\{\varphi_1,\ldots,\varphi_k\}$ of formulas (like an ``alternative set'' \cite{GoreW09,GoreW10}) with $k > 1$. The slowdown is not high as we check ``compatibility'' between nodes as soon as possible. A generalization of our method for this case is straightforward: we delete the edge $(u,v)$ and connect $u$ to nodes $v_0$, $v_1,\ldots,v_k$ (which are created when necessary) such that $v_0$ extends $v$ with $\{\varphi_1,\ldots,\varphi_k\}$ and, for each $1 \leq i \leq k$, $v_i$ is similar to $v$ but always disallows $\varphi_i$. 

Apart from~\cite{GBGI,dkns2011,GoreW09,GoreW10}, some other papers that are most related to ours are: Pratt's paper~\cite{Pratt80} on PDL, Demri's paper~\cite{Demri01} of REG, the paper~\cite{ddn-jlli05} by Demri and de Nivelle on \CReg, the paper~\cite{de-giacomo-massacci-converse-pdl} by De Giacomo and Massacci on CPDL. 
Also, one can translate the satisfiability problem of \CPDLreg into the satisfiability problem of modal $\mu$-calculus with converse or the emptiness problem of automata on infinite trees. The former one is often translated into the latter one (see Vardi's work~\cite{Vardi98}). Checking emptiness of an automaton on infinite trees can be reduced to checking who has a winning strategy in a graph of game, which is similar to ``and-or'' graphs used in the current paper. However, the construction of such a tree automaton itself is a weak point of the approach. As stated in~\cite{BaaderSattler01}, optimization techniques have not been adequately studied for theorem proving based on tree automata, and this is why the approach is not used in practice for theorem proving in description logics (which are closely related to modal logics).

The rest of this paper is structured as follows. In Section~\ref{section: prel} we give definitions for the logic \CPDLreg. In Section~\ref{sec:tableau} we present our tableau calculus for \CPDLreg. In Section~\ref{section: proc} we give an \EXPTIME tableau decision procedure for checking satisfiability of a set of formulas w.r.t.\ a set of global assumption in \CPDLreg. We conclude in Section~\ref{section: conc}. 
\ShortVersion{Due to the lack of space, proofs of our results are presented only in the long version~\cite{nCPDLreg-long} of the current paper. The technical report~\cite{nCPDLreg-long} contains also an illustrative example and more explanation.}


\section{Preliminaries}
\label{section: prel}

Let $\mindices_+$ be a finite set of symbols. For $\sigma \in \mindices_+$, we use $\cnv{\sigma}$ to denote a fresh symbol, called the {\em converse} of $\sigma$. Let $\mindices_- = \{\cnv{\sigma} \mid \sigma \in \mindices_+\}$ and $\mindices = \mindices_+ \cup \mindices_-$. We call $\mindices$ an {\em alphabet with converse}. For $\varrho = \cnv{\sigma} \in \mindices_-$, define $\cnv{\varrho} \defeq \sigma$.

A {\em context-free semi-Thue system} $S$ over $\mindices$ is a finite set of context-free production rules over alphabet $\mindices$. We say that $S$ is {\em symmetric} if, for every rule $\sigma \to \varrho_1\ldots \varrho_k$ of $S$, the rule $\cnv{\sigma} \to \cnv{\varrho}_k\ldots\cnv{\varrho}_1$ also belongs to~$S$.
A context-free semi-Thue system $S$ over $\mindices$ is called a {\em regular semi-Thue system} $S$ over $\mindices$ if, for every $\sigma \in \mindices$, the set of words derivable from $\sigma$ using the system is a regular language over $\mindices$.

\LongVersion{A context-free semi-Thue system is like a context-free grammar, but it has no designated start symbol and there is no distinction between terminal and non-terminal symbols.}

Similarly to~\cite{ddn-jlli05}, we assume that any regular semi-Thue system $S$ is given together with a mapping $\aut$ that associates each $\sigma \in \mindices$ with a finite automaton $\aut_\sigma$ recognizing words derivable from $\sigma$ using $S$. We assume that for $\sigma \in \mindices$, the word $\sigma$ is derivable from $\sigma$ by such a system. We call $\aut$ the {\em mapping specifying the finite automata of}~$S$.\footnote{Note that it is undecidable to check whether a context-free semi-Thue system is regular since it is undecidable whether the language generated by a linear grammar is regular~\cite{MS97}.}

A {\em finite automaton} $A$ over alphabet $\mindices$ is a tuple $\langle \mindices, Q, I, \delta, F\rangle$, where $Q$ is a finite set of states, $I \subseteq Q$ is the set of initial states, $\delta \subseteq Q \times \mindices \times Q$ is the transition relation, and $F \subseteq Q$ is the set of accepting states.
A {\em run} of $A$ on a word $\varrho_1 \ldots \varrho_k$ is a finite sequence of states $q_0, q_1, \ldots, q_k$ such that $q_0 \in I$ and $\delta(q_{i-1},\varrho_i,q_i)$ holds for every $1 \leq i \leq k$. It is an {\em accepting run} if $q_k \in F$.
We say that $A$ {\em accepts} a word $w$ if there exists an accepting run of $A$ on $w$. The set of words accepted by $A$ is denoted by~$\mL(A)$.

We use $\props$ to denote the set of {\em propositions} (i.e., atomic formulas) and use letters like $p$ to denote its elements. We call elements of $\mindices_+$ {\em atomic programs}, and call elements of $\mindices$ {\em simple programs}. We denote simple programs by letters like $\sigma$ and~$\varrho$.

{\em Formulas} and {\em programs} in the {\em base language} of \CPDLreg\ are defined respectively by the following BNF grammar rules:
\begin{eqnarray*}
\varphi & ::= &
    \top
    \mid \bot
    \mid p
    \mid \lnot \varphi
    \mid \varphi \to \varphi
    \mid \varphi \land \varphi
    \mid \varphi \lor \varphi
    \mid \lDmd{\alpha}\varphi
    \mid [\alpha]\varphi \\[0.5ex]
\alpha & ::= &
    \sigma
    \mid \alpha;\alpha
    \mid \alpha \cup \alpha
    \mid \alpha^*
    \mid \cnv{\alpha}
    \mid \varphi?
\end{eqnarray*}

We use letters like $\alpha$, $\beta$ to denote programs, and $\varphi$, $\psi$, $\xi$ to denote formulas.
Given binary relations $R_1$ and $R_2$, by $R_1 \circ R_2$ we denote their relational composition.

A {\em Kripke model} is a pair $M = \langle \Delta^M, \cdot^M\rangle$, where $\Delta^M$ is a set of {\em states}, and $\cdot^M$ is an interpretation function that maps each proposition $p$ to a subset $p^M$ of $\Delta^M$, and each atomic program $\sigma \in \mindices_+$ to a binary relation $\sigma^M$ on $\Delta^M$. The interpretation function is extended to interpret complex formulas and complex programs as follows:
\[
\begin{array}{l}
\top^M = \Delta^M,\;\;
\bot^M = \emptyset\\[0.4ex]
(\lnot\varphi)^M = \Delta^M \setminus \varphi^M,\;\;
(\varphi \to \psi)^M = (\lnot\varphi \lor \psi)^M\\[0.4ex]
(\varphi \land \psi)^M = \varphi^M \cap \psi^M,\;\;
(\varphi \lor \psi)^M = \varphi^M \cup \psi^M\\[0.4ex]
(\lDmd{\alpha}\varphi)^M = \{ x \in \Delta^M \mid
    \E y[\alpha^M(x,y) \land \varphi^M(y)] \}\\[0.4ex]
([\alpha]\varphi)^M = \{ x \in \Delta^M \mid
    \V y[\alpha^M(x,y) \to \varphi^M(y)] \}\\[1.5ex]
(\alpha;\beta)^M = \alpha^M \circ \beta^M
\\[0.4ex]
(\alpha \cup \beta)^M = \alpha^M \cup \beta^M,\ \
(\alpha^*)^M = (\alpha^M)^*\\[0.4ex]
(\cnv{\alpha})^M = (\alpha^M)^{-1} = \{(y,x) \mid (x,y) \in \alpha^M\}\\[0.4ex]
(\varphi?)^M = \{ (x,x) \mid \varphi^M(x) \}.
\end{array}
\]
We write $M,w \models \varphi$ to denote $w \in \varphi^M$. For a set $X$ of formulas, we write $M,w \models X$ to denote that $M,w \models \varphi$ for all $\varphi \in X$. If $M,w \models \varphi$ (respectively, $M,w \models X$), then we say that $M$ {\em satisfies} $\varphi$ (respectively, $X$) {\em at} $w$, and that $\varphi$ (respectively, $X$) is {\em satisfied at} $w$ in $M$. We say that $M$ {\em validates} $X$ (and $X$ is {\em valid in} $M$) if $M,w \models X$ for all $w \in \Delta^M$.

Note that the definition of $(\cnv{\sigma})^M$ is compatible with the assumption $\cnv{(\cnv{\sigma})} = \sigma$.

Let $S$ be a symmetric regular semi-Thue system over $\mindices$. The \CPDLreg logic corresponding to $S$, denoted by $\CPDL(S)$, is characterized by the class of Kripke models $M$ such that, for every rule $\sigma \to \varrho_1\ldots\varrho_k$ of $S$, we have that $\varrho_1^M\circ\cdots\circ \varrho_k^M \subseteq \sigma^M$. Such a structure is called an {\em $L$-model}, for $L = \CPDL(S)$.

Let $L$ be a \CPDLreg\ logic and $X$, $\Gamma$ be finite sets of formulas. We say that $X$ is {\em $L$-satisfiable} w.r.t.\ the set $\Gamma$ of global assumptions if there exists an $L$-model that validates $\Gamma$ and satisfies $X$ at some state.

\LongVersion{We say that two sets $X$ and $Y$ of formulas are {\em $L$-equivalent} if for every $L$-model $M$ and every state $w$ of $M$, $(M,w \models X)$ iff $(M,w \models Y)$.}

A formula or a program is in {\em negation-and-converse normal form} (NCNF) if it does not use the connective $\to$, uses the operator $\lnot$ only immediately before propositions, and uses the converse program constructor $\cnv{}$ only for atomic programs.

Every formula $\varphi$ (respectively, program $\alpha$) can be transformed to a formula $\varphi'$ (respectively, program $\alpha'$) in NCNF that is equivalent to $\varphi$ (respectively, $\alpha$) in the sense that for every Kripke model $M$, $\varphi^M = (\varphi')^M$ (respectively, $\alpha^M = (\alpha')^M$).
For example, the NCNF of formula $\lnot [\cnv{((\sigma_1 \cup \sigma_2);\sigma_3^*;(\lnot\lnot p)?)}] (q \lor \lnot r)$ is $\lDmd{p?;(\cnv{\sigma_3})^*;(\cnv{\sigma_1} \cup \cnv{\sigma_2})}(\lnot q \land r)$.
In this paper we assume that formulas and programs are represented in NCNF and write $\ovl{\varphi}$ to denote the NCNF
of~$\lnot\varphi$.

The {\em alphabet $\Sigma(\alpha)$} of a program $\alpha$ and the {\em regular language $\mL(\alpha)$} generated by $\alpha$ are specified as follows:\footnote{Note that $\Sigma(\alpha)$ contains not only simple programs but also expressions of the form~$(\varphi?)$, and a program $\alpha$ is a regular expression over its alphabet $\Sigma(\alpha)$.}
\[
\begin{array}{l@{\extracolsep{3em}}l}
\Sigma(\sigma) = \mindices & \mL(\sigma) =\{\sigma\}\\
\Sigma(\varphi?) = \{\varphi?\} & \mL(\varphi?) = \{\varphi?\}\\
\Sigma(\beta;\gamma) = \Sigma(\beta) \cup       \Sigma(\gamma) & \mL(\beta;\gamma) =
\mL(\beta).\mL(\gamma)\\
\Sigma(\beta \cup \gamma) = \Sigma(\beta)   \cup \Sigma(\gamma) & \mL(\beta \cup
\gamma) = \mL(\beta) \cup \mL(\gamma)\\
\Sigma(\beta^*) = \Sigma(\beta) & \mL(\beta^*) = (\mL(\beta))^*
\end{array}
\]
where for sets of words $M$ and $N$, $M.N \defeq \{\alpha\beta \mid \alpha \in M, \beta \in N\}$, $M^0 \defeq \{\varepsilon\}$ (where $\varepsilon$ denotes the empty word), $M^{n+1} \defeq M.M^n$ for $n \geq 0$, and $M^* \defeq \bigcup_{n \geq 0} M^n$.

We will use letters like $\omega$ to denote either a simple program from $\mindices$ or a~test (of the form $\varphi?$).
A word $\omega_1\ldots\omega_k \in \mL(\alpha)$ can be treated as the program $(\omega_1;\ldots;\omega_k)$, especially when it is interpreted in a~Kripke model. As a~finite automaton $A$ over alphabet $\Sigma(\alpha)$ corresponds to a~program (the regular expression recognizing the same language), it is interpreted in a~Kripke model as follows:
\begin{equation}
A^M = \bigcup \{\gamma^M \mid \gamma \in \mL(A)\}.\label{eq:sem-aut}
\end{equation}


\section{A Tableau Calculus for CPDL$_{\mbox{\footnotesize \bf reg}}$}
\label{sec:tableau}

From now on, let $S$ be a~symmetric regular semi-Thue system over $\mindices$, $\aut$ be the mapping specifying the finite automata of $S$, and $L$ be the \CPDLreg\ logic corresponding to~$S$. In this section we present a tableau calculus for checking $L$-satisfiability. 

For each program $\alpha$, let $\Aut_\alpha$ be a~finite automaton recognizing the regular language $\mL(\alpha)$.
For each program $\alpha \notin \mindices$, let $\aut_\alpha$ be a~finite automaton recognizing the language $\mL(\alpha')$, where $\alpha'$ is obtained from $\alpha$ by substituting each $\sigma \in \mindices$ not inside any test by a~regular expression representing $\mL(\aut_\sigma)$.

The automaton $\Aut_\alpha$ can be constructed from $\alpha$ in polynomial time, and $\aut_\alpha$ can be constructed in polynomial time in the length of $\alpha$ and the sizes of the automata $(\aut_\sigma)_{\sigma \in \mindices}$. Roughly speaking, $\aut_\alpha$ can be obtained from $\Aut_\alpha$ by simultaneously substituting each transition $(q_1,\sigma,q_2)$ by the automaton $\aut_\sigma$.

From now on, let $X$ and $\Gamma$ be finite sets of formulas in NCNF of the base language. For the tableau calculus defined here for checking $L$-satisfiability of $X$ w.r.t.\ the set $\Gamma$ of global assumptions we extend the base language with the auxiliary modal operators $\Box_\sigma$, $[A,q]$ and $\lDmd{A,q}$, where $\sigma \in \mindices$, $A$ is either $\aut_\alpha$ or $\Aut_\alpha$ for some program $\alpha$ occurring in $X$ or $\Gamma$ in the form $[\alpha]\varphi$ or $\lDmd{\alpha}\varphi$, and $q$ is a~state of $A$.
Here, $[A,q]$ and $\lDmd{A,q}$ stand respectively for $[(A,q)]$ and $\lDmd{(A,q)}$, where $(A,q)$ is the automaton that differs from $A$ only in that $q$ is its only initial state.
We call $[A,q]$ (respectively, $\lDmd{A,q}$) a~{\em universal} (respectively, {\em existential}) {\em automaton-modal operator}.

In the {\em extended language}, if $\varphi$ is a~formula, then $\Box_\sigma\varphi$, $[A,q]\varphi$ and $\lDmd{A,q}\varphi$ are also formulas. A formula $\Box_\sigma\varphi$ has the same semantics as $[\sigma]\varphi$, i.e. $(\Box_\sigma\varphi)^M \defeq ([\sigma]\varphi)^M$. The semantics of formulas $[A,q]\varphi$ and $\lDmd{A,q}\varphi$ are defined as usual, treating $(A,q)$ as a~program with semantics specified by~\eqref{eq:sem-aut}.
\LongVersion{

}
Despite that $\Box_\sigma\varphi$ has the same semantics as $[\sigma]\varphi$, the operator $\Box_\sigma$ behaves differently from $[\sigma]$ in our calculus.
\LongVersion{
Given a~Kripke model $M$ and a~state $w \in \Delta^M$, we have that $w \in ([A,q]\varphi)^M$ (respectively,  $w \in (\lDmd{A,q}\varphi)^M$) iff
\begin{quote}
$w_k \in \varphi^M$ for all (respectively, some) $w_k \in \Delta^M$ such that there exist a~word $\omega_1\ldots\omega_k$ (with $k \geq 0$) accepted by $(A,q)$ with $(w,w_k) \in (\omega_1;\ldots;\omega_k)^M$.
\end{quote}
The condition $(w,w_k) \in (\omega_1;\ldots;\omega_k)^M$ means there exist states $w_0 = w$, $w_1, \ldots, w_{k-1}$ of $M$ such that, for each $1 \leq i \leq k$, if $\omega_i \in \mindices$ then $(w_{i-1},w_i) \in \omega_i^M$, else $\omega_i = (\psi_i?)$ for some $\psi_i$ and $w_{i-1} = w_i$ and $w_i \in \psi_i^M$. Clearly, $\lDmd{A,q}$ is dual to $[A,q]$ in the sense that $\lDmd{A,q}\varphi \equiv \lnot[A,q]\lnot\varphi$ for any formula~$\varphi$.
}

We will use the following convention:
\begin{itemize}
\item given a~finite automaton $A$, we always assume that $A = \langle \Sigma_A, Q_A, I_A, \delta_A, F_A\rangle$
\item for $q \in Q_A$, we define $\delta_A(q) = \{(\omega,q') \mid (q,\omega,q') \in \delta_A\}$.
\end{itemize}

\LongVersion{
\begin{lemma} \label{lemma: prog-aut}
Let $M$ be a~Kripke model, $\alpha$ be a~program, $\varphi$ be a~formula, and $A$ be a~finite automaton over $\Sigma(\alpha)$. Then:
\begin{enumerate}
\item \label{prog-aut-1} if $I_A = \{q_1,\ldots,q_k\}$ then $A^M = (A,q_1)^M \cup \ldots \cup (A,q_k)^M$
\item \label{prog-aut-2} $\alpha^M = \Aut_\alpha^M$
\item \label{prog-aut-3} if $I_{\Aut_\alpha} = \{q_1,\ldots,q_k\}$ then $(\lDmd{\alpha}\varphi)^M = (\lDmd{\Aut_\alpha,q_1}\varphi
    \lor \ldots \lor \lDmd{\Aut_\alpha,q_k}\varphi)^M$
\item if $M$ is an $L$-model then
  \begin{enumerate}
  \item \label{prog-aut-4a} $\Aut_\alpha^M = \aut_\alpha^M$
  \item \label{prog-aut-4b} if $I_{\aut_\alpha} = \{q_1,\ldots,q_k\}$ then $([\alpha]\varphi)^M = ([\aut_\alpha,q_1]\varphi \land
      \ldots \land [\aut_\alpha,q_k]\varphi)^M$.
  \end{enumerate}
\end{enumerate}
\end{lemma}

\begin{proof}
The assertions~\ref{prog-aut-1} and~\ref{prog-aut-2} clearly hold. The assertion~\ref{prog-aut-3} follows from the
assertions~\ref{prog-aut-1} and~\ref{prog-aut-2}. For the assertion~\ref{prog-aut-4a}, just note that $M$ is an $L$-model and, for every
$\sigma \in \mindices$, $\aut_\sigma$ accepts $\sigma$. The assertion~\ref{prog-aut-4b} follows from the assertions \ref{prog-aut-1},
\ref{prog-aut-2}, and \ref{prog-aut-4a}.
\myEnd
\end{proof}
} 


In what follows we define tableaux as rooted ``and-or'' graphs. Such a graph is a tuple $G = (V,E,\nu)$, where $V$ is a set of nodes, $E \subseteq V \times V$ is a set of edges, $\nu \in V$ is the root, and each node $v \in V$ has a number of attributes. If there is an edge $(v,w) \in E$ then we call $v$ a {\em predecessor} of $w$, and call $w$ a {\em successor} of $v$. The set of all attributes of $v$ is called the {\em contents of $v$}. Attributes of tableau nodes are:
\begin{itemize}
\item $\Type(v) \in \{\State, \NonState\}$. If $\Type(v) = \State$ then we call $v$ a {\em state}, else we call $v$ a {\em non-state} (or an {\em internal} node). A state is never directly connected to a state.

\item $\Status(v) \in \{\Unexpanded$, $\Expanded$, $\Incomplete$, $\Unsat$, $\Sat\}$.

\item $\Label(v)$ is a finite set of formulas, called the label of $v$. 

\item $\RFormulas(v)$ is a finite set of formulas, called the set of reduced formulas of~$v$.

\item $\DFormulas(v)$ is a finite set of formulas, called the set of disallowed formulas of~$v$.

\item $\StatePred(v) \in V \cup \{\Null\}$ is called the state-predecessor of $v$. It is available only when $\Type(v) = \NonState$. 
If $v$ is a non-state and $G$ has no paths connecting a state to $v$ then $\StatePred(v) = \Null$. Otherwise, $G$ has exactly one state $u$ that is connected to $v$ via a path not containing any other states. In that case, $\StatePred(v) = u$.

\item $\CELabel(v)$ is a formula of the form $\lDmd{\sigma}\varphi$. It is called the coming edge label of $v$ and is available only when $v$ is a successor of a state. Informally, if $\CELabel(v) = \lDmd{\sigma}\varphi$ and $u$ is the predecessor of $v$ then $\lDmd{\sigma}\varphi \in \Label(u)$ and this formula is realized at $u$ by creating a transition to $v$ (transitions are done only for states).

\item $\AfterTransPred(v) \in V$ is called the after-transition-predecessor of~$v$. It is available only when $\Type(v) = \NonState$. If $v$ is a non-state and $v_0 = \StatePred(v) \neq \Null$) then there is exactly one successor $v_1$ of $v_0$ such that every path connecting $v_0$ to $v$ must go through $v_1$, and we have that $\AfterTransPred(v) = v_1$. If $v$ is a non-state and $\StatePred(v) = \Null$ then $\AfterTransPred(v)$ is set to the root $\nu$.

\item $\FormulaSC(v)$ is a formula called the formula suggested by converse. It is available only when $\Type(v) = \State$ and $\Status(v) = \Incomplete$. In that case, it means that some node $w$ with $\StatePred(w) = v$ wants $\varphi \in \Label(v) \cup \RFormulas(v)$ but it does not hold. The formula $\varphi$ is expected to be present at $v$ when, for $v_1 = \AfterTransPred(w)$, $\CELabel(v_1)$ is of the form $\lDmd{\sigma}\psi$ and $\Box_{\cnv{\sigma}}\varphi \in \Label(w)$.
\end{itemize}

We define 
\begin{eqnarray*}
\AFormulas(v) & = & \Label(v) \cup \RFormulas(v) \\
\LongVersion{\NDFormulas(v) & = & \{\ovl{\varphi} \mid \varphi \in \DFormulas(v)\} \\ }
\LongVersion{\FullLabel(v) & = & \AFormulas(v) \cup \NDFormulas(v) \\ }
\Kind(v) & = & 
  \left\{
  \begin{array}{l}
  \AndNode \textrm{ if } \Type(v) = \State\\
  \OrNode  \textrm{ if } \Type(v) = \NonState 
  \end{array}
  \right.\\
\BeforeFormingState(v) & = & \textrm{$v$ has a successor which is a state}\\
\AfterTrans(v) & = & (\AfterTransPred(v) = v)
\end{eqnarray*}
$\AFormulas(v)$ is called the available formulas of $v$. 
In an ``and-or'' graph, states play the role of {\em ``and''-nodes}, while non-states play the role of {\em ``or''-nodes}.\footnote{But this is not necessarily true for (``and-or'' graphs in) other logics.} 

By the {\em local graph} of a state $v$ we mean the subgraph of $G$ consisting of all the paths starting from $v$ and not containing any other states. 
Similarly, by the local graph of a non-state $v$ we mean the subgraph of $G$ consisting of all the path starting from $v$ and not containing any states.

We apply global state caching in the sense that if $v_1$ and $v_2$ are different states then $\Label(v_1) \neq \Label(v_2)$ or $\RFormulas(v_1) \neq \RFormulas(v_2)$ or $\DFormulas(v_1) \neq \DFormulas(v_2)$. 
If $v$ is a non-state such that $\AfterTrans(v)$ then we also apply global caching for the local graph of $v$. That is, if $w_1$ and $w_2$ are different nodes of the local graph of $v$ then $\Label(w_1) \neq \Label(w_2)$ or $\RFormulas(w_1) \neq \RFormulas(w_2)$ or $\DFormulas(w_1) \neq \DFormulas(w_2)$. 

Our calculus \cL\ for the \CPDLreg\ logic $L$ will be specified, amongst others, by a finite set of tableau rules, which are used to expand nodes of tableaux. A~{\em tableau rule} is specified with the following information:
\LongVersion{
\begin{itemize}
\item the kind of the rule: an ``and''-rule or an ``or''-rule
\item the conditions for applicability of the rule (if any)
\item the priority of the rule 
\item the number of successors of a node resulting from applying the rule to it, and the way to compute their contents.
\end{itemize}
}
\ShortVersion{
the kind of the rule (an ``and''-rule or an ``or''-rule), 
the conditions for applicability of the rule (if any), 
the priority of the rule, 
the number of successors of a node resulting from applying the rule to it, and the way to compute their contents.
}

Tableau rules are usually written downwards, with a set of formulas above the line as the {\em premise}, which represents the label of the node to which the rule is applied, and a number of sets of formulas below the line as the {\em (possible) conclusions}, which represent the labels of the successor nodes resulting from the application of the rule.
Possible conclusions of an ``or''-rule are separated by $\mid$, while conclusions of an ``and''-rule are separated by~$\&$. If a rule is a unary rule (i.e.\ a rule with only one possible conclusion) or an ``and''-rule then its conclusions are ``firm'' and we ignore the word ``possible''.
The meaning of an ``or''-rule is that if the premise is $L$-satisfiable w.r.t.\ $\Gamma$ then some of the possible conclusions are also $L$-satisfiable w.r.t.\ $\Gamma$, while the meaning of an ``and''-rule is that if the premise is $L$-satisfiable w.r.t.\ $\Gamma$ then all of the conclusions are also $L$-satisfiable w.r.t.\ $\Gamma$ (possibly in different states of the model under construction).

Such a representation gives only a part of the specification of the rules.

We use $Y$ to denote a set of formulas, write $Y,\varphi$ to denote $Y \cup \{\varphi\}$ and write $Y,\Gamma$ to denote $Y \cup \Gamma$.
Our {\em tableau calculus \cL} for the \CPDLreg\ logic $L$ w.r.t.\ the set $\Gamma$ of global assumptions consists of the rules which are partially specified in Table~\ref{table: cL} together with two special rules $\rFormingState$ and $\rConv$, which will be explained later. 

For any rule of \cL\ except $\rFormingState$ and $\rConv$, the distinguished formulas of the premise are called the {\em principal formulas} of the rule. The rules $\rFormingState$ and $\rConv$ have no principal formulas.

As usual, we assume that, for each rule of \cL described in Table~\ref{table: cL}, the principal formulas are not members of the set $Y$ which appears in the premise of the rule.

The rule $\rTrans$ is the only ``and''-rule and the only {\em transitional rule}. Instantiating this rule, for example, to the set $\{\lDmd{\sigma}p,\lDmd{\sigma}q,\Box_\sigma r\}$ as the premise and $\Gamma = \{s\}$ we get two conclusions: $\{p,r,s\}$ and $\{q,r,s\}$. 
Expanding a state $v$ in a tableau by the rule $\rTrans$, each successor $w_i$ of $v$ is created due to a corresponding principal formula $\lDmd{\sigma_i}\varphi_i$ of the rule and we have $\CELabel(w_i) = \lDmd{\sigma_i}\varphi_i$.

The other rules of \cL\ are ``or''-rules, which are also called {\em static rules}. 
The intuition behind distinguishing between static/transitional is that static rules do not change the state of the model under construction, while each conclusion of the transitional rule forces a~move to a~new state. The transitional rule is used to expand states of tableaux, while the static rules are used to expand non-states of tableaux. 

\begin{table}[t!]
\[
\begin{array}{|c|}
\hline
\ \\[-1.2ex]
\rAnd\; \fracc{Y, \varphi \land \psi}{Y, \varphi, \psi} \hspace{8.3em}
\rOr\; \fracc{Y, \varphi \lor \psi}{Y, \varphi \mid Y, \psi} \\
\ \\[-1.2ex]
\hline
\ \\[-1.2ex]
\rAutB\; \fracc{Y, [\alpha]\varphi}{Y, [\aut_\alpha,q_1]\varphi, \ldots, [\aut_\alpha,q_k]\varphi}\;\; \textrm{if } I_{\aut_\alpha} = \{q_1,\ldots,q_k\}\\
\ \\
\;\rAutD\; \fracc{Y, \lDmd{\alpha}\varphi}{Y, \lDmd{\Aut_\alpha,q_1}\varphi \mid \ldots \mid Y, \lDmd{\Aut_\alpha,q_k}\varphi}\;\; \textrm{if }
  \left\{
  \begin{array}{l}
  \alpha \notin \mindices,\; \alpha \textrm{ is not a~test},\\
  \textrm{and } I_{\Aut_\alpha} = \{q_1,\ldots,q_k\}
  \end{array}
  \right.\\
\ \\[-1.2ex]
\hline
\ \\[-1.2ex]
\textrm{if } \delta_A(q) = \{(\omega_1,q_1),\ldots,(\omega_k,q_k)\} \textrm{ and } q \notin F_A : \\
\ \\[-1.2ex]
\rBox\; \fracc{Y, [A,q]\varphi}{Y, \Box_{\omega_1}[A,q_1]\varphi, \ldots, \Box_{\omega_k}[A,q_k]\varphi}\\
\ \\[-1.2ex]
\rDmd\; \fracc{Y, \lDmd{A,q}\varphi}{Y, \lDmd{\omega_1}\lDmd{A,q_1}\varphi \mid \ldots \mid Y, \lDmd{\omega_k}\lDmd{A,q_k}\varphi}\\
\ \\[-1.2ex]
\hline
\ \\[-1.2ex]
\textrm{if } \delta_A(q) = \{(\omega_1,q_1),\ldots,(\omega_k,q_k)\} \textrm{ and } q \in F_A :\\
\ \\[-1.2ex]
\rBoxD\; \fracc{Y, [A,q]\varphi}{Y, \Box_{\omega_1}[A,q_1]\varphi, \ldots, \Box_{\omega_k}[A,q_k]\varphi, \varphi}\\
\ \\[-1.2ex]
\rDmdD\; \fracc{Y, \lDmd{A,q}\varphi}{Y, \lDmd{\omega_1}\lDmd{A,q_1}\varphi \mid \ldots \mid Y, \lDmd{\omega_k}\lDmd{A,q_k}\varphi \mid Y, \varphi}\\
\ \\[-1.2ex]
\hline
\ \\[-1.2ex]
\rBoxQm\; \fracc{Y, \Box_{(\psi?)}\varphi}{Y, \ovl{\psi} \mid Y, \varphi}
\hspace{8.3em}
\rDmdQm\; \fracc{Y, \lDmd{\psi?}\varphi}{Y, \psi, \varphi}\\
\ \\[-1.2ex]
\hline
\ \\[-1.2ex]
\,\;\rTrans\; \fracc{Y, \lDmd{\sigma_1}\varphi_1, \ldots, \lDmd{\sigma_k}\varphi_k}{Y_1,\varphi_1,\Gamma \;\&\; \ldots \;\&\; Y_k, \varphi_k, \Gamma}\;\;
\\ \mbox{}\\[-1.2ex]
\textrm{where }
  \left\{
  \begin{array}{l}
  \textrm{$k \geq 1$, $Y$ contains no formulas of the form $\lDmd{\sigma}\varphi$,}\\
  \textrm{and } Y_i = \{\psi : \Box_{\sigma_i}\psi \in Y\} \textrm{ for every } 1 \leq i \leq k
  \end{array}
  \right.\\[3ex]
\hline
\end{array}
\]

\medskip

\caption{Some rules of the tableau calculus \cL\ for a~\CPDLreg\ logic $L$.\label{table: cL}}
\VSpace{-3ex}
\end{table}

For any state $w$, every predecessor $v$ of $w$ is always a non-state. Such a node $v$ is expanded and connected to $w$ by the static rule $\rFormingState$. The nodes $v$ and $w$ correspond to the same state of the Kripke model under construction. In other words, the rule $\rFormingState$ ``transforms'' a non-state to a state. The idea is to separate internal nodes (i.e.~non-states) from states, which are globally cached. 
The rule $\rFormingState$ guarantees that, if $\BeforeFormingState(v)$ holds then $v$ has exactly one successor, which is a state.

Expanding a non-state $v$ of a tableau by a static rule $\rho \in \{\rAnd$, $\rOr$, $\rAutB$, $\rAutD$, $\rBox$, $\rBoxD$, $\rBoxQm\}$ which uses $\varphi$ as the principal formula we put $\varphi$ into the set $\RFormulas(w)$ of each successor $w$ of $v$ by setting $\RFormulas(w) := \RFormulas(v) \cup \{\varphi\}$. We use $\RFormulas(w)$ to disallow expanding $w$ by static rules which use a formula from $\RFormulas(w)$ as the principal formula (i.e. to block reducing the formulas from $\RFormulas(w)$ twice). If a non-state $v$ is expanded by a static rule $\rho \in \{\rDmd$, $\rDmdD$, $\rDmdQm$, $\rConv\}$ and $w$ is a successor of $v$ then we set $\RFormulas(w) := \RFormulas(v)$. Thus, we do not use the attribute $\RFormulas$ to disallow the rules $\rDmd$, $\rDmdD$, $\rDmdQm$ (in order to be able to fulfill eventualities of the form $\lDmd{A,q}\varphi$). If $v$ is expanded by the rule $\rTrans$ and $w$ is a successor of $v$ then we set $\RFormulas(w) := \emptyset$. 

Let $v_0, v_1, \ldots, v_k$ be a path of a tableau such that $k \geq 1$, $\Type(v_0) = \State$ and $\Type(v_i) = \NonState$ for $1 \leq i \leq k$. Suppose that $\CELabel(v_1) = \lDmd{\sigma}\psi$ and $\Box_{\cnv{\sigma}}\varphi \in \Label(v_k)$. If $v_0$ corresponds to a state $x$ of a Kripke model $M$, then all $v_1$, \ldots, $v_k$ correspond to a state $y$ of $M$ such that $(x,y) \in \sigma^M$. The formulas from $\Label(v_k)$ are supposed to be satisfied at $y$ in $M$, which causes $\varphi$ satisfied at $x$ in $M$. So, $\varphi$ is expected to belong to $\AFormulas(v_0)$ (i.e.\ we should realize $\varphi$ at $v_0$). What should be done in the case $\varphi \notin \AFormulas(v_0)\,$? We will not simply add $\varphi$ to $\Label(v_0)$, as $v_k$ is only one of possibly many ``or''-descendants of $v_0$, and adding $\varphi$ to $\Label(v_0)$ may affect the other ``or''-descendants of $v_0$ (which is not allowed). If $\varphi \in \DFormulas(v_0)$, which means $\varphi$ is disallowed in $v_0$, then $\Status(v_k)$ becomes $\Unsat$, which intuitively means that the ``combination'' of $v_0$ and $v_k$ is $L$-unsatisfiable w.r.t.\ $\Gamma$. In the other case, $\Status(v_0)$ becomes $\Incomplete$, the predecessors of $v_0$ will be re-expanded by the rule $\rConv$ either to have $\varphi$ (by adding $\varphi$ to the attribute $\Label$) or to disallow $\varphi$ (by adding $\varphi$ to the attribute $\DFormulas$). For details, see Steps~3-6 of procedure $\Apply$ given on page~\pageref{proc: Apply} (where $w$ plays the role of $v_0$). 

The priorities of the tableau rules are as follows (the bigger, the stronger): 
unary static rules except $\rFormingState$:~5; 
non-unary static rules:~4;
$\rFormingState$:~3;
$\rTrans$:~2; 
$\rConv$:~1.\footnote{The rule $\rConv$ is not chosen like the others but only invoked at another part of the algorithm (namely Procedure~$\UpdateStatus$).}

The conditions for applying a rule $\rho \neq \rConv$ to a node $v$ are as follows: 
\begin{itemize}
\item the rule has $\Label(v)$ as the premise
\item all the conditions accompanying with $\rho$ in Table~\ref{table: cL} are satisfied
\item if $\rho = \rTrans$ then $\Type(v) = \State$
\item if $\rho \neq \rTrans$ then $\Type(v) = \NonState$ and 
  \begin{itemize}
  \item if $\rho \in \{\rAnd, \rOr, \rAutB, \rAutD, \rBox, \rBoxD, \rBoxQm\}$ then the principal formula of $\rho$ does not belong to $\RFormulas(v)$ 
  \item no static rule with a higher priority is applicable to~$v$.
  \end{itemize}
\end{itemize}


\begin{figure*}[t!]
\begin{function}[H]
\caption{NewSucc($v, type, ceLabel, label, rFmls, dFmls$)\label{proc: NewSucc}}
\GlobalData{a rooted graph $(V,E,\nu)$.}
\Purpose{create a new successor for $v$.}
\LinesNumberedHidden
create a new node $w$,\ \ 
$V := V \cup \{w\}$,\ \ 
\lIf{$v \neq \Null$}{$E := E \cup \{(v,w)\}$}\;

$\Type(w) := type$,\ 
$\Status(w) := \Unexpanded$\; 
$\Label(w) := label$,\ 
$\RFormulas(w) := rFmls$,\ 
$\DFormulas(w) := dFmls$\; 

\If{$type = \NonState$}{
   \uIf{$v = \Null$ or $\Type(v) = \State$}{
	$\StatePred(w) := v$,\ \ 
	$\AfterTransPred(w) := w$,\ \ 
	$\CELabel(w) := ceLabel$
   }
   \lElse{
	$\StatePred(w) := \StatePred(v)$,\ \ 
	$\AfterTransPred(w) := \AfterTransPred(v)$
   }
}

\Return{w}
\end{function}

\begin{function}[H]
\caption{FindProxy($type, root, label, rFmls, dFmls$)\label{proc: FindProxy}}
\GlobalData{a rooted graph $(V,E,\nu)$.}
\LinesNumberedHidden

\lIf{$type = \State$}{$W := V$}
\lElse{$W := $ the nodes of the local graph of $root$}\;

\lIf{there exists $v \in W$ such that $\Type(v) = type$ and $\Label(v) = label$ and $\RFormulas(v) = rFmls$ and $\DFormulas(v) = dFmls$}{\Return $v$\\}
\lElse{\Return $\Null$}
\end{function}

\begin{function}[H]
\caption{ConToSucc($v, type, ceLabel, label, rFmls, dFmls$)\label{proc: ConToSucc}}
\GlobalData{a rooted graph $(V,E,\nu)$.}
\Purpose{connect $v$ to a successor, which is created if necessary.}

\lIf{$type = \State$}{$root := \Null$}
\lElse{$root := \AfterTransPred(v)$}

$w := \FindProxy(type, root, label, rFmls, dFmls)$\;
\lIf{$w \neq \Null$}{$E := E \cup \{(v,w)\}$\\}
\lElse{$w := \NewSucc(v, type, ceLabel, label, rFmls, dFmls)$}\;

\Return{w}
\end{function}

\begin{function}[H]
\caption{TUnsat($v$)\label{proc: TUnsat}}
\LinesNumberedHidden

\Return{$\bot \in \Label(v)$ or there exists $\{\varphi,\ovl{\varphi}\} \subseteq \Label(v)$}
\end{function}

\begin{function}[H]
\caption{TSat($v$)\label{proc: TSat}}
\LinesNumberedHidden

\Return{no rule except $\rConv$ is applicable to $v$}
\end{function}

\begin{function}[H]
\caption{ToExpand()\label{proc: ToExpand}}
\GlobalData{a rooted graph $(V,E,\nu)$.}
\lIf{there exists $v \in V$ with $\Status(v) = \Unexpanded$}{\Return \texttt{/* any such */} $v$\\}
\lElse{\Return $\Null$}\tcp*[r]{various ``search strategies'' can be applied here}
\end{function}

\end{figure*}


\begin{procedure}[t!]
\caption{Apply($\rho, v$)\label{proc: Apply}}
\GlobalData{a rooted graph $(V,E,\nu)$.}
\Input{a rule $\rho$ and a node $v \in V$ s.t. if $\rho \neq \rConv$ then $\Status(v) = \Unexpanded$ else $\Status(v) = \Expanded$ and $\BeforeFormingState(v)$.}
\Purpose{applying the tableau rule $\rho$ to the node $v$.}

\BlankLine

  \uIf{$\rho = \rFormingState$}{
     $\ConToSucc(v,\State,\Null,\Label(v),\RFormulas(v), \DFormulas(v))$\;
  }
  \uElseIf{$\rho = \rConv$}{
     let $w$ be the only successor of $v$,\ \ 
     $E := E \setminus \{(v,w)\}$,\ \ 
     $\varphi := \FormulaSC(w)$\;

     $\ConToSucc(v,\NonState,\Null, \Label(v) \cup \{\varphi\}, \RFormulas(v), \DFormulas(v))$\;

     $\ConToSucc(v,\NonState,\Null,\Label(v), \RFormulas(v), \DFormulas(v) \cup \{\varphi\})$\;

  }
  \uElseIf{$\rho = \rTrans$}{
     let $Y_1$, \ldots, $Y_k$ be the possible conclusions of the rule\;
     let $\psi_1$, \ldots, $\psi_k$ be the corresponding principal formulas\;
     \lForEach{$1 \leq i \leq k$}{$\NewSucc(v,\NonState,\psi_i,Y_i,\emptyset, \emptyset)$}
  }
  \Else{
     let $Y_1$, \ldots, $Y_k$ be the possible conclusions of the rule\;
     \lIf{$\rho \in \{\rDmd$, $\rDmdD$, $\rDmdQm$\}}{$Z := \RFormulas(v)$\\}
     \lElse{$Z := \RFormulas(v) \cup \{\textrm{the principal formula of $\rho$}\}$}\;
     \lForEach{$1 \leq i \leq k$}{$\ConToSucc(v, \NonState, \Null, Y_i, Z, \DFormulas(v))$}
  }

\ForEach{successor $w$ of $v$ with $\Status(w) \notin \{\Incomplete,\Unsat,\Sat\}$}{
   \lIf{$\TUnsat(w)$}{$\Status(w) := \Unsat$}
   \Else{
      \If{$\Type(w) = \NonState$}{
	$u_0 := \StatePred(w)$,\ \ 
	$u_1 := \AfterTransPred(w)$\;
	\If{there exist $\Box_{\cnv{\sigma}}\varphi \in \Label(w)$, $\CELabel(u_1) = \lDmd{\sigma}\psi$, $\varphi \notin \AFormulas(u_0)$}{
	   \lIf{$\varphi \in \DFormulas(u_0)$}{$\Status(w) := \Unsat$\\}
	   \Else{$\Status(u_0) := \Incomplete$,\ \ 
		$\FormulaSC(u_0) := \varphi$\;
		$\PropagateStatus(u_0)$,\ \ 
		\Return 
	   }
	}
      }

      \BlankLine
      \lIf{$\Status(w) \neq \Unsat$ and $\TSat(w)$}{$\Status(w) := \Sat$}
   }
}

$\Status(v) := \Expanded$,\ \ 
$\UpdateStatus(v)$\;
\lIf{$\Status(v) \in \{\Incomplete,\Sat,\Unsat\}$}{$\PropagateStatus(v)$}
\end{procedure}


\begin{figure*}[t!]
\begin{function}[H]
\caption{Tableau($X, \Gamma$)\label{proc: Tableau}}
\Input{finite sets $X$ and $\Gamma$ of concepts in NCNF of the base language.}
\GlobalData{a rooted graph $(V,E,\nu)$.}
\BlankLine

$\nu := \NewSucc(\Null,\NonState,\Null,X \cup \Gamma, \emptyset, \emptyset)$\;
\lIf{$\TUnsat(\nu)$}{$\Status(\nu) := \Unsat$\\}
\lElseIf{$\TSat(\nu)$}{$\Status(\nu) := \Sat$}\;

\BlankLine
\While{$(v := \ToExpand()) \neq \Null$}{
choose a tableau rule $\rho$ different from $\rConv$ and applicable to $v$\;
$\Apply(\rho, v)$\tcp*[l]{defined on page \pageref{proc: Apply}}
\ForEach{$w \in V$ such that $\Status(w) = \Incomplete$}
  {delete the local graph of $w$ except the node $w$}
} 

\Return $(V,E,\nu)$
\end{function}

\medskip

\begin{procedure}[H]
\caption{UpdateStatus($v$)\label{proc: UpdateStatus}}
\GlobalData{a rooted graph $(V,E,\nu)$.}
\Input{a node $v \in V$ with $\Status(v) = \Expanded$.}
\BlankLine

\uIf{$\Kind(v) = \OrNode$}{
   \lIf{some successors of $v$ have status $\Sat$}{$\Status(v) := \Sat$\\}
   \lElseIf{all successors of $v$ have status $\Unsat$}{$\Status(v) := \Unsat$\\}
   \lElseIf{a successor of $v$ has status $\Incomplete$}{$\Apply(\rConv,v)$}
}
\Else(\tcp*[h]{$\Kind(v) = \AndNode$}){
   \lIf{all successors of $v$ have status $\Sat$}{$\Status(v) := \Sat$\\}
   \lElseIf{some successors of $v$ have status $\Unsat$}{$\Status(v) := \Unsat$}
}
\end{procedure}


\medskip

\begin{procedure}[H]
\caption{PropagateStatus($v$)\label{proc: PropagateStatus}}
\GlobalData{a rooted graph $(V,E,\nu)$.}
\Input{a node $v \in V$ with $\Status(v) \in \{\Incomplete,\Unsat,\Sat\}$.}
\BlankLine

\ForEach{predecessor $u$ of $v$ with $\Status(u) = \Expanded$}{
   $\UpdateStatus(u)$\;
   \lIf{$\Status(u) \in \{\Incomplete,\Unsat,\Sat\}$}{$\PropagateStatus(u)$}
}

\end{procedure}
\end{figure*}


Application of a tableau rule $\rho$ to a node $v$ is specified by procedure $\Apply(\rho, v)$ given on page \pageref{proc: Apply}. Auxiliary functions are defined on page~\pageref{proc: NewSucc}. Procedures used for updating and propagating statuses of nodes are defined on page~\pageref{proc: Tableau}. 
The main function $\Tableau(X,\Gamma)$ is also defined on page~\pageref{proc: Tableau}. It returns a rooted ``and-or'' graph called a {\em \cL-tableau} for $(X,\Gamma)$.


\LongVersion{
\begin{example}\label{ex:andorgraph}
Consider the regular grammar logic with converse $L$ that corresponds to the following semi-Thue system over alphabet $\{\sigma,\varrho,\cnv{\sigma},\cnv{\varrho}\}$:
\[ \{\varrho \to \cnv{\sigma}\varrho,\ \varrho \to \sigma,\ \cnv{\varrho} \to \cnv{\varrho}\sigma,\ \cnv{\varrho} \to \cnv{\sigma}\}. \]
The set of words derivable from $\varrho$ is characterized by $(\cnv{\sigma})^*(\sigma + \varrho)$.
Let
\[ \aut_\varrho = \langle \{\sigma,\varrho,\cnv{\sigma},\cnv{\varrho}\}, \{0,1\}, \{0\}, \{(0,\cnv{\sigma},0), (0,\sigma,1), (0,\varrho,1)\}, \{1\}\rangle.\]

In Figures~\ref{fig: example1} and \ref{fig: example1-II} we give an ``and-or'' graph for  $(\{\lDmd{\sigma}(p \land [\varrho]\lnot p)\}, \emptyset)$
w.r.t.\ \cL.
The nodes are numbered when created and are expanded using~DFS (depth-first search). At the end the root receives status $\Unsat$. Therefore, by Theorem~\ref{theorem: s-c}, $\lDmd{\sigma}(p \land [\varrho]\lnot p)$ is $L$-unsatisfiable.
\myEnd
\end{example}

\newcommand{\myhline}{\\[0.4ex] \hline \\[-1.7ex]}
\newcommand{\ehline}{\ \\[-0.7ex] \hline \ \\[-0.7ex]}

\begin{figure}
\begin{center}
\begin{tabular}{c}
\begin{scriptsize}
\begin{tabular}{c@{\extracolsep{10em}}c}
(a) & (b) \\
\\
\xymatrix{
*+[F]{\begin{tabular}{c}
	(1): \rFormingState
	\myhline
	$\lDmd{\sigma}(p \land [\varrho]\lnot p)$
      \end{tabular}}
\ar@{->}[d] 
\\
*+[F=]{\begin{tabular}{c}
	(2): \rTrans
	\myhline
	$\lDmd{\sigma}(p \land [\varrho]\lnot p)$
      \end{tabular}}
\ar@{->}[d] 
\\
*+[F]{\begin{tabular}{c}
	(3): $\rAnd$
	\myhline
	$p \land [\varrho]\lnot p$
      \end{tabular}}
\ar@{->}[d]
\\
*+[F]{\begin{tabular}{c}
	(4): $\rAutB$
	\myhline
	$p, [\varrho]\lnot p$
      \end{tabular}}
\ar@{->}[d]
\\
*+[F]{\begin{tabular}{c}
	(5): $\rBox$
	\myhline
	$p, [\aut_\varrho,0]\lnot p$
      \end{tabular}}
\ar@{->}[d]
\\
*+[F]{\begin{tabular}{c}
	(6): $\rBox$
	\myhline
	$p, \Box_{\cnv{\sigma}}[\aut_\varrho,0]\lnot p$,\\
	$\Box_\sigma[\aut_\varrho,1]\lnot p$ 
      \end{tabular}}
}
&
\xymatrix{
*+[F]{\begin{tabular}{c}
	(1): \rFormingState
	\myhline
	$\lDmd{\sigma}(p \land [\varrho]\lnot p)$
      \end{tabular}}
\ar@{->}[d] 
\\
*+[F=]{\begin{tabular}{c}
	(2): \rTrans
	\myhline
	$\lDmd{\sigma}(p \land [\varrho]\lnot p)$\\
	$\Incomplete$\\
	$\FormulaSC: [\aut_\varrho,0]\lnot p$
      \end{tabular}}
\ar@{->}[d] 
\\
*+[F]{\begin{tabular}{c}
	(3): $\rAnd$
	\myhline
	$p \land [\varrho]\lnot p$
      \end{tabular}}
\ar@{->}[d]
\\
*+[F]{\begin{tabular}{c}
	(4): $\rAutB$
	\myhline
	$p, [\varrho]\lnot p$
      \end{tabular}}
\ar@{->}[d]
\\
*+[F]{\begin{tabular}{c}
	(5): $\rBox$
	\myhline
	$p, [\aut_\varrho,0]\lnot p$
      \end{tabular}}
\ar@{->}[d]
\\
*+[F]{\begin{tabular}{c}
	(6): $\rBox$
	\myhline
	$p, \Box_{\cnv{\sigma}}[\aut_\varrho,0]\lnot p$,\\
	$\Box_\sigma[\aut_\varrho,1]\lnot p$ 
      \end{tabular}}
}
\end{tabular}
\end{scriptsize}
\end{tabular}
\end{center}
\caption{\label{fig: example1}An illustration for Example~\ref{ex:andorgraph}: part I. The graph (a) is the ``and-or'' graph constructed until checking ``compatibility'' of the node (6) w.r.t. to the node (1). 
In each node, we display the name of the rule expanding the node and the formulas of the label of the node. The attribute $\DFormulas$ of each of the displayed nodes is an empty set. The node (2) is the only state. As an example, we have $\StatePred((6)) = (2)$, $\AfterTransPred((6)) = (3)$ and $\CELabel((3)) = \lDmd{\sigma}(p \land [\varrho]\lnot p)$. Since $\Box_{\cnv{\sigma}}[\aut_\varrho,0]\lnot p \in \Label((6))$, $\Status((2))$ is set to $\Incomplete$ and $\FormulaSC((2))$ is set to $[\aut_\varrho,0]\lnot p$. This results in the graph~(b). The construction is then continued by applying the rule $\rConv$ to (1) and deleting the nodes (3)-(6). See Figure~\ref{fig: example1-II} for the continuation.} 
\end{figure}
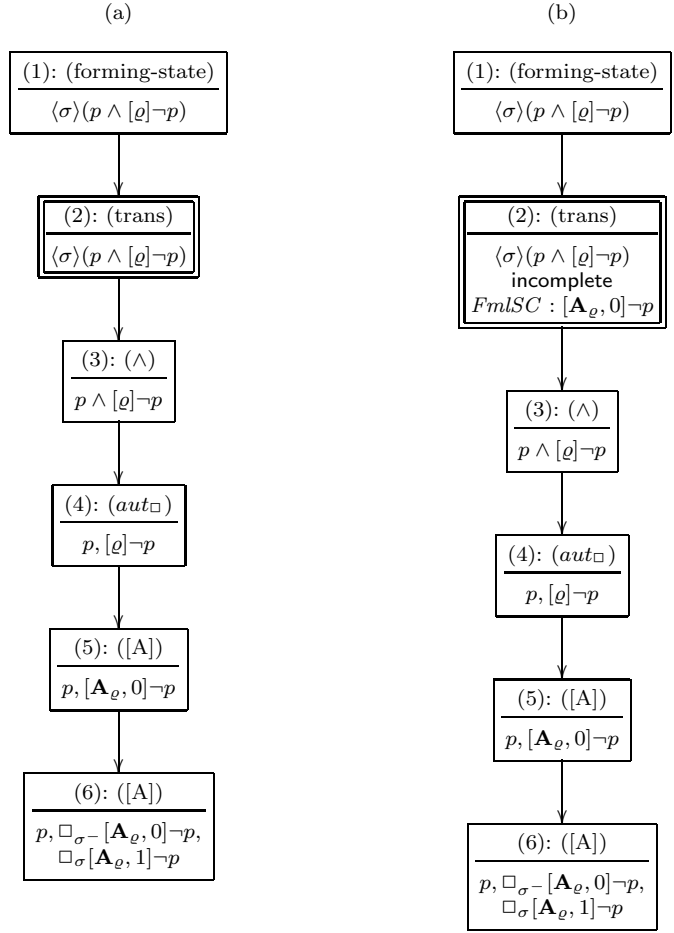


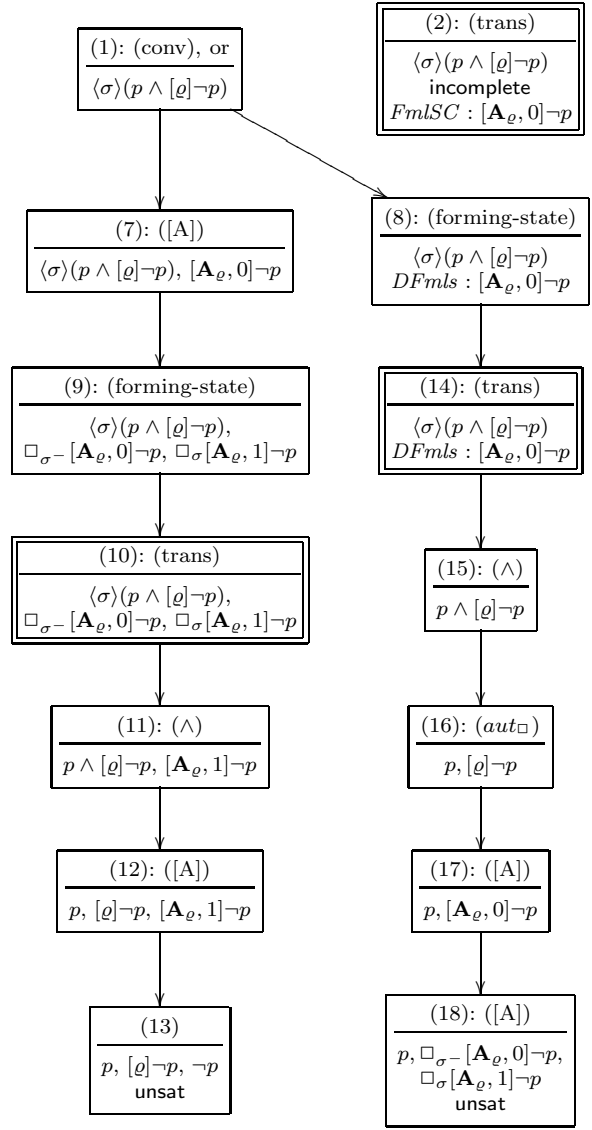
\begin{figure}
\begin{center}
\begin{tabular}{c}
\begin{scriptsize}
\xymatrix{
*+[F]{\begin{tabular}{c}
	(1): \rConv, or
	\myhline
	$\lDmd{\sigma}(p \land [\varrho]\lnot p)$
      \end{tabular}}
\ar@{->}[d] 
\ar@{->}[rd] 
&
*+[F=]{\begin{tabular}{c}
	(2): \rTrans
	\myhline
	$\lDmd{\sigma}(p \land [\varrho]\lnot p)$\\
	$\Incomplete$\\
	$\FormulaSC: [\aut_\varrho,0]\lnot p$
      \end{tabular}}
\\
*+[F]{\begin{tabular}{c}
	(7): $\rBox$
	\myhline
	$\lDmd{\sigma}(p \land [\varrho]\lnot p)$, 
	$[\aut_\varrho,0]\lnot p$
      \end{tabular}}
\ar@{->}[d] 
&
*+[F]{\begin{tabular}{c}
	(8): \rFormingState
	\myhline
	$\lDmd{\sigma}(p \land [\varrho]\lnot p)$\\
	$\DFormulas: [\aut_\varrho,0]\lnot p$
      \end{tabular}}
\ar@{->}[d] 
\\
*+[F]{\begin{tabular}{c}
	(9): \rFormingState
	\myhline
	$\lDmd{\sigma}(p \land [\varrho]\lnot p)$,\\ 
	$\Box_{\cnv{\sigma}}[\aut_\varrho,0]\lnot p$,
	$\Box_\sigma[\aut_\varrho,1]\lnot p$
      \end{tabular}}
\ar@{->}[d] 
&
*+[F=]{\begin{tabular}{c}
	(14): \rTrans
	\myhline
	$\lDmd{\sigma}(p \land [\varrho]\lnot p)$\\
	$\DFormulas: [\aut_\varrho,0]\lnot p$
      \end{tabular}}
\ar@{->}[d] 
\\
*+[F=]{\begin{tabular}{c}
	(10): \rTrans
	\myhline
	$\lDmd{\sigma}(p \land [\varrho]\lnot p)$,\\ 
	$\Box_{\cnv{\sigma}}[\aut_\varrho,0]\lnot p$,
	$\Box_\sigma[\aut_\varrho,1]\lnot p$
      \end{tabular}}
\ar@{->}[d] 
&
*+[F]{\begin{tabular}{c}
	(15): $\rAnd$
	\myhline
	$p \land [\varrho]\lnot p$
      \end{tabular}}
\ar@{->}[d]
\\
*+[F]{\begin{tabular}{c}
	(11): $\rAnd$ 
	\myhline
	$p \land [\varrho]\lnot p$,
	$[\aut_\varrho,1]\lnot p$
      \end{tabular}}
\ar@{->}[d] 
&
*+[F]{\begin{tabular}{c}
	(16): $\rAutB$
	\myhline
	$p, [\varrho]\lnot p$
      \end{tabular}}
\ar@{->}[d]
\\
*+[F]{\begin{tabular}{c}
	(12): $\rBox$
	\myhline
	$p$, $[\varrho]\lnot p$,
	$[\aut_\varrho,1]\lnot p$
      \end{tabular}}
\ar@{->}[d] 
&
*+[F]{\begin{tabular}{c}
	(17): $\rBox$
	\myhline
	$p, [\aut_\varrho,0]\lnot p$
      \end{tabular}}
\ar@{->}[d]
\\
*+[F]{\begin{tabular}{c}
	(13)
	\myhline
	$p$, $[\varrho]\lnot p$,
	$\lnot p$\\
	$\Unsat$
      \end{tabular}}
&
*+[F]{\begin{tabular}{c}
	(18): $\rBox$
	\myhline
	$p, \Box_{\cnv{\sigma}}[\aut_\varrho,0]\lnot p$,\\
	$\Box_\sigma[\aut_\varrho,1]\lnot p$\\ 
	$\Unsat$
      \end{tabular}}
}
\end{scriptsize}
\end{tabular}
\end{center}
\caption{\label{fig: example1-II}An illustration for Example~\ref{ex:andorgraph}: part II. This is a fully expanded ``and-or'' graph for $(\{\lDmd{\sigma}(p \land [\varrho]\lnot p)\}, \emptyset)$ w.r.t. \cL. The node (1) is re-expanded by the rule $\rConv$. As in the part~I, in each node we display the name of the rule expanding the node and the formulas of the label of the node. We display also the attribute $\DFormulas$ of the nodes (8) and (9). This attribute of any other node is an empty set. The nodes (2), (10) and (14) are the only states. The node (13) receives status $\Unsat$ because $\{p,\lnot p\} \subset \Label((13))$. After that the nodes (12)-(7) receive status $\Unsat$ in subsequent steps. The node (18) receives status $\Unsat$ because $[\aut_\varrho,0]\lnot p \in \DFormulas((14))$. After that the nodes (17)-(14), (8), (1) receive status $\Unsat$ in subsequent steps.} 
\end{figure}
} 

\ShortVersion{See the long version~\cite{nCPDLreg-long} of this paper for an example of ``and-or'' graph.}


Observe that:
\begin{itemize}
\item An application of the rule $\rConv$ to a node $v$ may cause a sequence of other applications of this rule to ``ancestor nodes'' of $v$ and may put formulas far back of the tableau.
\item If the logic $L$ is essentially a \CReg logic, then no formulas of the form $\lDmd{A,q}\varphi$ occur in tableaux (and we do not have to check ``global consistency'').
\end{itemize}


\LongVersion{
Define the {\em length} of a~formula $\varphi$ to be the number
of symbols occurring in~$\varphi$. For example, the length of
$\lDmd{\aut_\sigma,q}\psi$ is the length of $\psi$ plus 5,
treating $\aut_\sigma$ as a~symbol. Define the {\em size} of
a~finite set of formulas to be the length of the conjunction of
its formulas.
Define the {\em size} of a~finite automaton $\langle
\Sigma,Q,I,\delta,F\rangle$ to be $|Q| + |I| + |\delta| + |F|$.

For a set $Y$ of formulas, the set of {\em basic subformulas of $Y$}, denoted by $\bsf(Y)$, consists of all formulas $\varphi$ and $\ovl{\varphi}$ of the base language such that either $\varphi \in Y$ or $\varphi$ is a subformula of some formula of $Y$. The set $\clsL(Y)$ is defined to be the union of $\bsf(Y)$ and the following two sets:
\begin{itemize}
\item $\{[\aut_\alpha,q]\varphi$, $\Box_\omega[\aut_\alpha,q]\varphi \mid$ $[\alpha]\varphi \in \bsf(Y)$, $q \in Q_{\aut_\alpha}$ and
    $\omega \in \Sigma_{\aut_\alpha}\}$

\item $\{\lDmd{\Aut_\alpha,q}\varphi$, $\lDmd{\omega}\lDmd{\Aut_\alpha,q}\varphi \mid$ $\lDmd{\alpha}\varphi \in \bsf(Y)$, $\alpha \notin \mindices$, $\alpha$ is not a test, $q \in Q_{\Aut_\alpha}$ and $\omega \in \Sigma_{\Aut_\alpha}\}$.
\end{itemize}

\begin{lemma} \label{lemma: properties of tableaux}
Let $h = |\Sigma|$ and let 
$k$ be the sum of the sizes of the automata $\aut_\sigma$ (for $\sigma \in \Sigma$), $l$ be the size of $X \cup \Gamma$, and $n$ be the size of $\clsL(X \cup \Gamma)$. 
Let $G = (V,E,\nu)$ be a \cL-tableau for $(X,\Gamma)$. 
Then $n$ is polynomial in $k.h.l$ and, for every $v \in V\,$:
\begin{enumerate}
\item \label{item: HHSKA 1}
The sets $\Label(v)$, $\RFormulas(v)$, $\DFormulas(v)$ and $\NDFormulas(v)$ contain only formulas from $\clsL(X \cup \Gamma)$.

\item \label{item: HHSKA 2}
$\RFormulas(v)$ does not contain formulas of the form $\lDmd{A,q}\varphi$ or $\lDmd{\psi?}\varphi$.

\item \label{item: HHSKA 3}
If $v$ is a state then $\Label(v) \cap \RFormulas(v) = \emptyset$ and $\Label(v)$ does not contain formulas of the form $\varphi \land \psi$, $\varphi \lor \psi$, $[\alpha]\varphi$, $[A,q]\varphi$, $\Box_{(\psi?)}\varphi$, $\lDmd{A,q}\varphi$ or $\lDmd{\beta}\varphi$ with $\beta \notin \mindices$.

\item \label{item: HHSKA 4}
  \begin{enumerate}
  \item $\Label(v)$ is $L$-equivalent to $\Label(v) \cup \RFormulas(v)$
  \item $\Label(v) \setminus \RFormulas(v)$ is $L$-equivalent to $\Label(v) \cup \RFormulas(v)$.
  \end{enumerate}
\end{enumerate}
Furthermore, the graph $G$ has no more than $2^{O(n)}$ nodes and can be constructed in $2^{O(n)}$ steps.
\end{lemma}

\begin{proof}
Clearly, $n$ is polynomial in $k.h.l$. 
The assertions \ref{item: HHSKA 1}-\ref{item: HHSKA 3} should be clear. The assertion~\ref{item: HHSKA 4} can be proved by induction in a straightforward way. 
Since global state caching is used, by assertion~\ref{item: HHSKA 1}, $G$ has no more than $2^{O(n)}$ states. Similarly, since global caching is used for the local graphs of non-states, the local graph of each non-state has no more than $2^{O(n)}$ nodes. Hence, $G$ has no more than $2^{O(n)}$ nodes. 
Each node of the graph may be re-expanded at most once, using the rule $\rConv$. Expansion of a node can be done in polynomial time in the size of the graph. Hence the graph can be constructed in $2^{O(n)}$ steps. 
\myEnd
\end{proof}
} 


A {\em marking} of a \cL-tableau $G$ is a~subgraph $G_m$ of $G$ such that:
\begin{itemize}
\item the root of $G$ is the root of $G_m$ \item if $v$ is a~node of $G_m$ and is an ``or''-node of $G$ then at least one edge $(v,w)$ of
    $G$ is an edge of~$G_m$
\item if $v$ is a~node of $G_m$ and is an ``and''-node of $G$ then every edge $(v,w)$ of $G$ is an edge of $G_m$
\item if $(v,w)$ is an edge of $G_m$ then $v$ and $w$ are nodes of $G_m$.
\end{itemize}

Let $G$ be a \cL-tableau for $(X,\Gamma)$, $G_m$ be a marking of $G$, $v$ be a node of~$G_m$, and $\lDmd{A,q}\varphi$ be a formula of $\Label(v)$. A {\em trace} of $\lDmd{A,q}\varphi$ in $G_m$ starting from $v$ is a sequence $(v_0,\varphi_0)$, \ldots, $(v_k,\varphi_k)$ such that:
\begin{itemize}
\item $v_0 = v$ and $\varphi_0 = \lDmd{A,q}\varphi$
\item for every $1 \leq i \leq k$, $(v_{i-1},v_i)$ is an edge of $G_m$ 
\item for every $1 \leq i \leq k$, $\varphi_i$ is a formula of $\Label(v_i)$ such that 
  \begin{itemize}
  \item if $\varphi_{i-1}$ is not a principal formula of the tableau rule expanding $v_{i-1}$ then the rule must be a static rule and $\varphi_i = \varphi_{i-1}$
  \item else if the rule is $\rDmd$ or $\rDmdD$ then $\varphi_{i-1}$ is of the form $\lDmd{A,q'}\varphi$ and $\varphi_i$ is the formula obtained from~$\varphi_{i-1}$
  \item else if the rule is $\rDmdQm$ then $\varphi_{i-1}$ is of the form $\lDmd{\psi?}\lDmd{A,q'}\varphi$ and $\varphi_i = \lDmd{A,q'}\varphi$
  \item else the rule is $\rTrans$, $\varphi_{i-1}$ is of the form $\lDmd{\sigma}\lDmd{A,q'}\varphi$ and is the coming edge label of $v_i$, and $\varphi_i = \lDmd{A,q'}\varphi$.
  \end{itemize}
\end{itemize}

A trace $(v_0,\varphi_0)$, \ldots, $(v_k,\varphi_k)$ of $\lDmd{A,q}\varphi$ in a marking $G_m$ is called a {\em $\Dmd$-realization in $G_m$ for $\lDmd{A,q}\varphi$ at $v_0$} if $\varphi_k = \varphi$.


A marking $G_m$ of a \cL-tableau $G$ is {\em consistent} if:
\begin{description}
\item[local consistency:] $G_m$ does not contain nodes with status $\Unsat$; and
\item[global consistency:] for every node $v$ of $G_m$, every formula $\lDmd{A,q}\varphi$ of $\Label(v)$ has a $\Dmd$-realization (starting from $v$) in $G_m$.
\end{description}

\begin{theorem}[Soundness and Completeness]
\label{theorem: s-c} Let $S$ be a~symmetric regular semi-Thue system over $\Sigma$, $\aut$ be the mapping specifying the finite automata of $S$, and $L$ be the \CPDLreg\ logic corresponding to $S$. Let $X$ and $\Gamma$ be finite sets of formulas in NCNF of the base language, and $G$ be a \cL-tableau for $(X,\Gamma)$. Then $X$ is $L$-satisfiable w.r.t.\ the set $\Gamma$ of global assumptions iff $G$ has a consistent marking.
\end{theorem}

\ShortVersion{See the long version~\cite{nCPDLreg-long} of this paper for the proof of this theorem.}


\LongVersion{
\section{Proofs of Soundness and Completeness}
\label{section: proofs}

Let $G$ be a \cL-tableau for $(X,\Gamma)$. For each $v \in V$ with $\Status(v) \in \{\Incomplete$, $\Unsat$, $\Sat\}$, let $\DSTimeStamp(v)$ be the step number at which $\Status(v)$ is changed to its final value (i.e.\ determined to be $\Incomplete$, $\Unsat$ or $\Sat$). $\DSTimeStamp$ stands for ``determined-status timestamp''. 

\subsection{Soundness}

\begin{lemma} \label{lemma: GSDHE}
Let $G = (V,E,\nu)$ be a \cL-tableau for $(X,\Gamma)$. Then no node with status $\Incomplete$ is reachable from $\nu$. 
\end{lemma}

\begin{proof}
By Steps~4, 25, 26 of procedure $\Apply$ and Step~4 of procedure $\UpdateStatus$.
\myEnd
\end{proof}

\begin{lemma} \label{lemma: a-fl}
If a non-state $v$ has $w_1$, \ldots, $w_k$ as all the successors then, for every $L$-model $M$ and every $x \in \Delta^M$, we have that $M,x \models \FullLabel(v)$ iff there exists $1 \leq i \leq k$ such that $M,x \models \FullLabel(w_i)$.
\myEnd
\end{lemma}

The proof of this lemma is straightforward.

\begin{lemma} \label{lemma: SHQWD}
Let $G = (V,E,\nu)$ be a \cL-tableau for $(X,\Gamma)$. Let $v \in V$ be a node with $\Status(v) = \Unsat$. Then:
\begin{itemize}
\item case $\Type(v) = \State$ : $\FullLabel(v)$ is $L$-unsatisfiable w.r.t.~$\Gamma$
\item case $\Type(v) = \NonState$ and $\StatePred(v) = \Null$ : $\FullLabel(v)$ is $L$-unsatisfiable w.r.t.~$\Gamma$

\item case $\Type(v) = \NonState$ and $v_0 = \StatePred(v) \neq \Null$ : there are no $L$-model $M$ and $x,y \in \Delta^M$ such that $M$ validates $\Gamma$, $(x,y) \in \sigma^M$, $M,x \models \FullLabel(v_0)$ and $M,y \models \FullLabel(v)$, where $v_1 = \AfterTransPred(v)$ and $\CELabel(v_1)$ is of the form $\lDmd{\sigma}\psi$.
\end{itemize}
\end{lemma}

\begin{proof}
We prove this by induction on $\DSTimeStamp(v)$. 

If $\bot \in \Label(v)$ or there exists $\{\varphi,\ovl{\varphi}\} \subseteq \Label(v)$ then $\FullLabel(v)$ is clearly $L$-unsatisfiable w.r.t.~$\Gamma$. So, in the rest of this proof, we exclude this case. 

Consider the case when $\Type(v) = \State$.

We have that $\Kind(v) = \AndNode$. 
There exists a successor $w$ of $v$ with $\Status(w) = \Unsat$ and $\DSTimeStamp(w) < \DSTimeStamp(v)$. By the inductive assumption, $\FullLabel(w)$ is $L$-unsatisfiable w.r.t.~$\Gamma$. Since $\DFormulas(w) = \emptyset$, by assertion~\ref{item: HHSKA 4} of Lemma~\ref{lemma: properties of tableaux}, it follows that $\Label(w)$ is $L$-unsatisfiable w.r.t.~$\Gamma$. Hence $\Label(v)$ and $\FullLabel(v)$ are $L$-unsatisfiable w.r.t.~$\Gamma$. 

Consider the case when $\Type(v) = \NonState$ and $\StatePred(v) = \Null$.

We have that $\Kind(v) = \OrNode$.
Let $w_1, \ldots, w_k$ be all the successors of $v$. 
We must have that, for all $1 \leq i \leq k$, $\Status(w_i) = \Unsat$ and $\DSTimeStamp(w_i) < \DSTimeStamp(v)$. By the inductive assumption, for all $1 \leq i \leq k$, $\FullLabel(w_i)$ is $L$-unsatisfiable w.r.t.~$\Gamma$. 
By Lemma~\ref{lemma: a-fl}, $\FullLabel(v)$ is $L$-unsatisfiable w.r.t.~$\Gamma$.

Consider the case when $\Type(v) = \NonState$ and $v_0 = \StatePred(v) \neq \Null$.
\begin{itemize}
\item Case when $v$ has a successor $w$ being a state: The node $v$ must be expanded by the rule $\rFormingState$. 
Since $w$ is the only successor of $v$, it must be that $\Status(w) = \Unsat$ and $\DSTimeStamp(w) < \DSTimeStamp(v)$. By the inductive assumption, $\FullLabel(w)$ is $L$-unsatisfiable w.r.t.~$\Gamma$. Hence $\FullLabel(v) = \FullLabel(w)$ is also $L$-unsatisfiable w.r.t.~$\Gamma$. 

\item Case when $v$ has a successor and all the successors $w_1, \ldots, w_k$ of $v$ are non-states. 

  \begin{itemize}
  \item If $\Status(v)$ was set to $\Unsat$ by Step~20 of procedure $\Apply$ (with $w = v$) then, for $v_1 = \AfterTransPred(v)$, $\CELabel(v_1)$ is of the form $\lDmd{\sigma}\psi$, $\Label(v)$ contains a formula of the form $\Box_{\cnv{\sigma}}\varphi$ and $\varphi \notin \AFormulas(v_0)$, $\varphi \in \DFormulas(v_0)$. For the sake of contradiction, suppose that there exist an $L$-model $M$ and $x,y \in \Delta^M$ such that $M$ validates $\Gamma$, $(x,y) \in \sigma^M$, $M,x \models \FullLabel(v_0)$ and $M,y \models \FullLabel(v)$.
Since $\Box_{\cnv{\sigma}}\varphi \in \Label(v)$, it follows that $M,y \models \Box_{\cnv{\sigma}}\varphi$, and hence $M,x \models \varphi$. This contradicts the facts that $\varphi \in \DFormulas(v_0)$ and $M,x \models \FullLabel(v_0)$.
  \item Consider the remaining case. It must be that, for all $1 \leq i \leq k$, $\Status(w_i) = \Unsat$ and $\DSTimeStamp(w_i) < \DSTimeStamp(v)$. By the inductive assumption, for all $1 \leq i \leq k$, $\FullLabel(w_i)$ is $L$-unsatisfiable w.r.t.~$\Gamma$. By Lemma~\ref{lemma: a-fl}, $\FullLabel(v)$ is $L$-unsatisfiable w.r.t.~$\Gamma$.
\myEnd
  \end{itemize}
\end{itemize}
\end{proof}

\begin{lemma}[Soundness] \label{lemma: soundness CPDLreg}
Let $G = (V,E,\nu)$ be a \cL-tableau for $(X,\Gamma)$. Suppose that $X$ is $L$-satisfiable w.r.t.\ the set $\Gamma$ of global assumptions. Then $G$ has a~consistent marking.
\end{lemma}

\begin{proof}
Since $\Type(\nu) = \NonState$, $\Label(\nu) = X \cup \Gamma$ and $\DFormulas(\nu) = \emptyset$, we have that $\FullLabel(\nu)$ is $L$-satisfiable w.r.t.~$\Gamma$. Let $M$ be an $L$-model validating $\Gamma$ and let $\tau$ be a state of $M$ such that $M,\tau \models \FullLabel(\nu)$.
 
Observe that, for any $v \in V$ and any $x \in \Delta^M$, if $M,x \models \FullLabel(v)$ then: 
\begin{itemize}
\item if $v$ is a non-state then $v$ has a successor $w$ such that $M,x \models \FullLabel(w)$ (by Lemma~\ref{lemma: a-fl})
\item if $v$ is a state then, for every successor $w$ of $v$ with $\CELabel(w)$ of the form $\lDmd{\sigma}\psi$, there exists $y \in \Delta^M$ such that $(x,y) \in \sigma^M$ and $M,y \models \FullLabel(w)$.\footnote{Note that $\CELabel(w) \in \Label(v)$.}
\end{itemize}
Therefore, starting from $\nu \in V$ and $\tau \in \Delta^M$ it is straightforward to construct a marking $G_m$ of $G$ together with a map $g$ that associates each edge $(v,w)$ of $G_m$ with a pair $(x,y) \in \Delta^M \times \Delta^M$ such that:
\begin{itemize}
\item if $v$ is a non-state then $x = y$
\item if $v$ is a state and $\CELabel(w)$ is of the form $\lDmd{\sigma}\psi$ then $(x,y) \in \sigma^M$
\item $M,x \models \FullLabel(v)$ and $M,y \models \FullLabel(w)$.
\end{itemize}

By Lemma~\ref{lemma: SHQWD}, for every node $v$ of $G_m$, $\Status(v) \neq \Unsat$. Therefore, $G_m$ satisfies the local consistency property. We now show that $G_m$ satisfies the global consistency property.
Let $v_0$ be a node of $G_m$ and let $\lDmd{A,q}\varphi \in \Label(v_0)$. We show that $\lDmd{A,q}\varphi$ has a~$\Dmd$-realization in $G_m$ starting from $v_0$. By the map $g$, there exists $u \in \Delta^M$ such that $M,u \models \FullLabel(v_0)$. Thus $M,u \models \lDmd{A,q}\varphi$ and therefore, 
\begin{equation}\label{eq:star}
\parbox{10cm}{there exist a~word $\gamma = \omega_1\ldots\omega_k \in \mL((A,q))$ with $k \geq 0$, an accepting run $q_0 = q$, $q_1$, \ldots, $q_k$ of $(A,q)$ on $\gamma$, and states $u_0 = u$, $u_1$, \ldots, $u_k$ of $M$ such that $M,u_k \models \varphi$ and, for $1 \leq i \leq k$, if $\omega_i \in \mindices$ then $(u_{i-1},u_i) \in \omega_i^M$, else $\omega_i$ is of the form $(\psi_i?)$ and $u_{i-1} = u_i \in \psi_i^M$.}
\end{equation}
%

We construct a~$\Dmd$-realization $(v_0,\varphi_0),\ldots,(v_h,\varphi_h)$ in $G_m$ for $\lDmd{A,q}\varphi$ at $v_0$ and a~map $f :
\{0,\ldots,h\} \to \{0,\ldots,k\}$ such that $f(0) = 0$, $f(h) = k$, and for every $0 \leq i < h$, if $f(i) = j$ then $f(i+1)$ is either
$j$ or $j+1$.
We maintain the following invariants for $0 \leq i \leq h\,$:
\begin{align}
&\mbox{--\;\;the sequence
    $(v_0,\varphi_0),\ldots,(v_i,\varphi_i)$ is a~trace of
    $\lDmd{A,q}\varphi$ in $G_m$}\label{eq:inv-a}\\
&\mbox{--\;\;$\FullLabel(v_i)$ is satisfied at the state
    $u_{f(i)}$ of $M$}\label{eq:inv-b}\\
&\mbox{--\;\;if $f(i) = j < k$ then
    $\varphi_i = \lDmd{A, q_j}\varphi$ or
    $\varphi_i = \lDmd{\omega_{j+1}}\lDmd{A,
    q_{j+1}}\varphi$}\label{eq:inv-c}\\
&\mbox{--\;\;if $f(i) = k$ then $\varphi_i =
    \lDmd{A, q_k}\varphi$ or $\varphi_i =
    \varphi$.}\label{eq:inv-d}
\end{align}
With $\varphi_0 = \lDmd{A,q_0}\varphi$ and $f(0) = 0$, the invariants clearly hold for $i = 0$.

\noindent Set $i := 0$. While $\varphi_i \neq \varphi$ do:
\begin{enumerate}
\item Set $j := f(i)$. 

\item Case $v_i$ is expanded using a static rule $\rho$ and $\varphi_i$ is the principal formula: 
  \begin{enumerate}
  \item Case $j < k$ : Since $\rho$ is a static rule, by the invariant~\eqref{eq:inv-c}, we must have $\varphi_i = \lDmd{A, q_j}\varphi$ or $\varphi_i = \lDmd{\omega_{j+1}}\lDmd{A, q_{j+1}}\varphi$ with $\omega_{j+1} = (\psi_{j+1}?)$.

      \smallskip

      Consider the case $\varphi_i = \lDmd{A, q_j}\varphi$. The applied rule $\rho$ is thus either $\rDmd$ or $\rDmdD$. Let $\varphi_{i+1} = \lDmd{\omega_{j+1}}\lDmd{A,q_{j+1}}\varphi$ and let $v_{i+1}$ be the successor of $v_i$ with $\varphi_{i+1}$ replacing $\varphi_i$. By~\eqref{eq:star}, $\varphi_{i+1}$ is satisfied at $u_j$ in $M$, and hence, by the invariant~\eqref{eq:inv-b}, $\FullLabel(v_{i+1})$ is satisfied at $u_j$ in $M$. Let $f(i+1) = j$ and set $i := i+1$. Clearly, the invariants still hold.

      \smallskip

      Now consider the case $\varphi_i = \lDmd{\omega_{j+1}}\lDmd{A, q_{j+1}}\varphi$ with $\omega_{j+1} = (\psi_{j+1}?)$. The applied rule $\rho$ is thus $\rDmdQm$. Let $\varphi_{i+1} = \lDmd{A,q_{j+1}}\varphi$ and let $v_{i+1}$ be the only successor of~$v_i$. By~\eqref{eq:star}, both $\psi_{j+1}$ and $\lDmd{A,q_{j+1}}\varphi$ are satisfied at $u_j$ in $M$, and hence, by the invariant~\eqref{eq:inv-b}, $\FullLabel(v_{i+1})$ is satisfied at $u_j = u_{j+1}$ in $M$. Let $f(i+1) = j+1$ and set $i := i+1$. Clearly, the invariants still hold.

  \item Case $j = k$ : Since $\varphi_i \neq \varphi$, by the invariant~\eqref{eq:inv-d}, we have that $\varphi_i = \lDmd{A, q_k}\varphi$. Hence $\rho = \rDmdD$ (since $q_k \in F_A$). Let $\varphi_{i+1} = \varphi$ and let $v_{i+1}$ be the successor of $v_i$ with $\varphi_{i+1}$ replacing $\varphi_i$. By~\eqref{eq:star}, $\varphi_{i+1}$ is satisfied at $u_k$ in $M$, and hence, by the invariant~\eqref{eq:inv-b}, $\FullLabel(v_{i+1})$ is satisfied at $u_k$ in $M$. Let $f(i+1) = k$ and set $i := i+1$. Clearly, the invariants still hold.
  \end{enumerate}

\item Case $v_i$ is expanded using a static rule $\rho$ and either $\rho$ does not have principal formulas or $\varphi_i$ is not the principal formula: 
  \begin{itemize}
  \item Case $\rho$ does not have principal formulas or the principal formula is not of the form $\lDmd{A',q'}\varphi'$: Let $\varphi_{i+1} = \varphi_i$ and $f(i+1) = f(i) = j$. 
  Let $v_{i+1}$ be the successor of $v_i$ such that $(v_i,v_{i+1})$ is an edge of $G_m$ and $\FullLabel(v_{i+1})$ is satisfied at the state $u_{f(i+1)}$ of $M$. Such a node $v_{i+1}$ exists because $\FullLabel(v_i)$ is satisfied at the state $u_{f(i)} = u_j = u_{f(i+1)}$ of $M$. 
  By setting $i := i+1$, all the invariants~\eqref{eq:inv-a}-\eqref{eq:inv-d} still hold.

  \item Case $\rho$ has the principal formula of the form $\lDmd{A',q'}\varphi'$: During a sequence of applications of static rules between two applications of the transitional rule, proceed as for realizing $\lDmd{A',q'}\varphi'$ in $G_m$ (like for the current $\Dmd$-realization of $\lDmd{A,q}\varphi$ in $G_m$ at $v_0$). This decides how to choose $v_{i+1}$ and has effects on terminating the trace (to obtain a $\Dmd$-realization for $\lDmd{A,q}\varphi$ in $G_m$ at $v_0$). We also choose $\varphi_{i+1} = \varphi_i$ and $f(i+1) = f(i) = j$. By setting $i := i+1$, all the invariants~\eqref{eq:inv-a}-\eqref{eq:inv-d} still hold.
  \end{itemize}

\item Case $v_i$ is expanded using the transitional rule:\\
Since $v_i$ is a state and $\varphi_i \neq \varphi$, by the invariants~\eqref{eq:inv-c} and \eqref{eq:inv-d}, we must have that $\varphi_i = \lDmd{\omega_{j+1}}\lDmd{A,q_{j+1}}\varphi$ with $\omega_{j+1} \in \mindices$. Let $v_{i+1}$ be the successor of $v_i$ with $\CELabel(v_{i+1}) = \varphi_i$. Let $\varphi_{i+1} = \lDmd{A,q_{j+1}}\varphi$ and $f(i+1) = j+1$. Clearly, the invariant~\eqref{eq:inv-a} holds for $i+1$. By~\eqref{eq:star}, $\varphi_{i+1}$ is satisfied at the state $u_{j+1}$ of $M$. By the invariant~\eqref{eq:inv-b}, the other formulas of $\Label(v_{i+1})$ are also satisfied at the state $u_{j+1}$ of $M$. That is, the invariant~\eqref{eq:inv-b} holds for $i+1$. Clearly, the invariants~\eqref{eq:inv-c} and~\eqref{eq:inv-d} remain true after increasing $i$ by 1. So, by setting $i := i+1$, all the invariants~\eqref{eq:inv-a}-\eqref{eq:inv-d} still hold.
\end{enumerate}

We now show that the loop terminates. 
Observe that any sequence of applications of static rules that contribute to the trace $(v_0,\varphi_0),\ldots,(v_i,\varphi_i)$ of $\lDmd{A,q}\varphi$ in $G_m$ eventually ends because:
\begin{itemize}
\item each formula not of the forms $\lDmd{A',q'}\varphi'$ and $\lDmd{\psi'?}\varphi'$ may be reduced at most once
\item each formula of the form $\lDmd{\psi'?}\varphi'$ is reduced to $\psi'$ and $\varphi'$ 
\item each formula of the form $\lDmd{A',q'}\varphi'$ is reduced according to some $\Dmd$-realization. 
\end{itemize}

That is, sooner or later either $\varphi_i = \varphi$ or $v_i$ is a node that is expanded by the transitional rule. In the second case, if $f(i) = j$ then $f(i+1) = j+1$. As the image of $f$ is $\{0,\ldots,k\}$, the construction of the trace must end at some step (with $\varphi_i = \varphi$) and we obtain a~$\Dmd$-realization in $G_m$ for $\lDmd{A,q}\varphi$ at $v_0$. 
\myEnd
\end{proof}


\subsection{Model Graphs}

We will prove completeness of \cL\ via model graphs~\cite{Rautenberg83,Gore99,nguyen01B5SL,NguyenSzalas-CADE-22,NguyenSzalas-KSE09}.

\begin{Definition}
A {\em model graph} (also known as a {\em Hintikka structure}) is a tuple $\langle W,R,H\rangle$, where $W$ is a set of nodes, $R$ is a mapping that associates each $\sigma \in \mindices$ with a binary relation $R_\sigma$ on $W$, and $H$ is a mapping that associates each node of $W$ with a set of formulas.
\end{Definition}

We use model graphs merely as data structures, but we are interested in consistent and saturated model graphs as defined below.
Model graphs differ from ``and-or'' graphs in that a~model graph contains only ``and''-nodes and its edges are labeled by simple programs.
Roughly speaking, given an ``and-or'' graph $G$ with a~consistent marking $G_m$, to construct a~model graph one can stick together the nodes
in a~``saturation path'' of a~node of $G_m$ to create a~node for the model graph. Details will be given later.

A trace of a~formula $\lDmd{A,q}\varphi$ at a~node in a~model graph is defined analogously as for the case of ``and-or'' graphs, as stated
in the following definition.

\begin{Definition}
Given a~model graph $M = \langle W, R, H\rangle$ and a~node $v \in W$, a~{\em trace} of a~formula $\lDmd{A,q}\varphi \in H(v)$ (starting
from $v$) is a~sequence $(v_0,\varphi_0)$, \ldots, $(v_k,\varphi_k)$ such that:
\begin{itemize}
\item $v_0 = v$ and $\varphi_0 = \lDmd{A,q}\varphi$,
\item for every $1 \leq i \leq k$, $\varphi_i \in H(v_i)$,
\item for every $1 \leq i \leq k$, if $v_i = v_{i-1}$ then:
  \begin{itemize}
  \item $\varphi_{i-1}$ is of the form $\lDmd{A,q'}\varphi$ and $\varphi_i = \lDmd{\omega}\lDmd{A,q''}\varphi$ for some $\omega$
      and $q''$ such that $(q',\omega,q'') \in \delta_A$, or
  \item $\varphi_{i-1}$ is of the form $\lDmd{A,q'}\varphi$ with $q' \in F_A$, $\varphi_i = \varphi$, and $i = k$, or
  \item $\varphi_{i-1}$ is of the form $\lDmd{\psi?}\lDmd{A,q'}\varphi$ and $\varphi_i = \lDmd{A,q'}\varphi$
  \end{itemize}
\item for every $1 \leq i \leq k$, if $v_i \neq v_{i-1}$ then $\varphi_{i-1}$ is of the form $\lDmd{\sigma}\lDmd{A,q'}\varphi$, $\sigma
    \in \mindices$, $\varphi_i = \lDmd{A,q'}\varphi$,
   and $(v_{i-1},v_i) \in R_\sigma$.
\end{itemize}

A trace $(v_0,\varphi_0)$, \ldots, $(v_k,\varphi_k)$ of $\lDmd{A,q}\varphi$ in a~model graph $M$ is called a~{\em $\Dmd$-realization} for
$\lDmd{A,q}\varphi$ at $v_0$ if $\varphi_k = \varphi$.
\end{Definition}

Similarly as for markings of ``and-or'' graphs, we define the consistency of a~model graph as follows.

\begin{definition}\em
A~model graph $M = \langle W, R, H\rangle$ is {\em consistent} if:
\begin{itemize}
\item {\em local consistency:} for every $v \in W$, $H(v)$ contains neither $\bot$ nor a~{\em clashing pair} of the form $p$ and $\lnot p\,$; and
\item {\em global consistency:} for every $v \in W$, every formula $\lDmd{A,q}\varphi$ of $H(v)$ has a~$\Dmd$-realization (at~$v$).
\myEnd
\end{itemize}
\end{definition}

\begin{definition}\em
A model graph $M = \langle W,R,H\rangle$ is said to be {\em \cL-saturated} if the following conditions hold for every $v \in W$ and every $\varphi \in H(v)$:
  \begin{itemize}
  \item if $\varphi = \psi \land \xi$ then $\{\psi,\xi\} \subset H(v)$
  \item if $\varphi = \psi \lor \xi$ then $\psi \in H(v)$ or $\xi \in H(v)$

  \item if $\varphi\! =\! [\alpha]\psi$ and $I_{\aut_\alpha}\! =\! \{q_1,\ldots,q_k\}$ then
      $\{[\aut_\sigma,q_1]\psi,\ldots,[\aut_\sigma,q_k]\psi\}\! \subset\! H(v)$

  \item if $\varphi = [A,q]\psi$ and $\delta_A(q) = \{(\omega_1,q_1)$, \ldots, $(\omega_k,q_k)\}$ then\\ \mbox{\hspace{1.5cm}}
      $\{\Box_{\omega_1}[A,q_1]\psi, \ldots, \Box_{\omega_k}[A,q_k]\psi\} \subset H(v)$

  \item if $\varphi = [A,q]\psi$ and $q \in F_A$ then $\psi \in H(v)$

  \item if $\varphi = \Box_{(\xi?)}\psi$ then $\ovl{\xi} \in H(v)$ or $\psi \in H(v)$

  \item if $\varphi = \lDmd{\alpha}\psi$, $\alpha \notin \mindices$, $\alpha$ is not a~test and $I_{\Aut_\alpha} =
      \{q_1,\ldots,q_k\}$ then\\ \mbox{\hspace{1.5cm}} one of the formulas $\lDmd{\Aut_\alpha,q_1}\psi$, \ldots,
      $\lDmd{\Aut_\alpha,q_k}\psi$ belongs to $H(v)$

  \item if $\varphi = \lDmd{\xi?}\psi$ then $\xi \in H(v)$ \footnote{The condition $\psi \in H(v)$ is taken care of by the global
      consistency.}

  \item if $\varphi = \lDmd{\sigma}\psi$ then there exists $w$ such that $R_\sigma(v,w)$ and $\psi \in H(w)$

  \item if $\varphi = \Box_\sigma\psi$ and $R_\sigma(v,w)$ holds then $\psi \in H(w)$
  \item if $\varphi = \Box_\sigma\psi$ and $R_{\cnv{\sigma}}(u,v)$ holds then $\psi \in H(u)$.\myEnd
  \end{itemize}
\end{definition}

\begin{definition}\em
Given a~model graph $M = \langle W,R,H\rangle$, the {\em $L$-model corresponding to $M$} is the Kripke model $M'$ such that
$\Delta^{M'} = W$, $p^{M'} = \{w \in W \mid p \in H(w)\}$ for $p \in \props$, $\varrho^{M'} = R'_\varrho\,$ for $\varrho \in
\mindices^+$, where $R'_\sigma\,$ for $\sigma \in \mindices$ are the smallest binary relations on $W$ such that:
  \begin{itemize}
  \item $R_\sigma \subseteq R'_\sigma$ and $R'_{\cnv{\sigma}} = (R'_\sigma)^{-1}$ for every $\sigma \in \mindices$, and
  \item if $\sigma \to \varrho_1\ldots\varrho_k \in S$, where $S$ is the symmetric regular semi-Thue system of $L$, then
      $R'_{\varrho_1}\circ\cdots\circ R'_{\varrho_k} \subseteq R'_\sigma$.\footnote{Note that the symmetry of $S$ is essential here.}\myEnd
  \end{itemize}
\end{definition}
The smallest binary relations mentioned in the above definition exist because:
\begin{itemize}
\item $R'_{\cnv{\sigma}} = (R'_\sigma)^{-1}$ iff $(R'_{\cnv{\sigma}})^{-1} \subseteq R'_\sigma$ and $(R'_\sigma)^{-1} \subseteq R'_{\cnv{\sigma}}$
\item an expression of the form $R^{-1} \subseteq R'$ is equivalent to $\V x, y (R(y,x) \to R'(x,y))$
\item an expression of the form $R_1 \circ\cdots\circ R_h \subseteq R'$ is equivalent to 
\[ \V x_0,\ldots,x_h\,(R_1(x_0,x_1) \land\ldots\land R_h(x_{h-1},x_h) \to R'(x_0,x_h)) \]
\item a set of first-order program clauses like the ones listed above is called a Datalog program and always has the smallest model.
\end{itemize} 

Define the NCNF of $\lnot\Box_\sigma\varphi$ to be $\lDmd{\sigma}\ovl{\varphi}$. Recall that the NCNF of $\lnot[\alpha]\varphi$,
$\lnot\lDmd{\alpha}\varphi$, $\lnot[A,q]\varphi$, $\lnot\lDmd{A,q}\varphi$ are $\lDmd{\alpha}\ovl{\varphi}$, $[\alpha]\ovl{\varphi}$,
$\lDmd{A,q}\ovl{\varphi}$, $[A,q]\ovl{\varphi}$, respectively.

\begin{lemma} \label{lemma: gen clash}
Let $M = \langle W,R,H\rangle$ be a~consistent and \cL-saturated model graph and let $M'$ be the $L$-model corresponding to $M$.
Then, for any $w \in W$, if $M',w \models \varphi$ or $\varphi \in H(w)$ then $\ovl{\varphi} \notin H(w)$.
\end{lemma}

\begin{proof}
By induction on the structure of $\varphi$, using the global consistency.
\myEnd
\end{proof}

\begin{lemma} \label{lemma: aut-Aut}
Let $\alpha$ be a program, $q \in Q_{\aut_\alpha}$, and $\gamma \in \mL((\aut_\alpha,q))$. If the rule $\sigma \to \varrho_1\ldots\varrho_k$ belongs to $S$ then replacing any occurrence of $\sigma$ in $\gamma$ by $\varrho_1\ldots\varrho_k$ results in $\gamma' \in \mL((\aut_\alpha,q))$.
\end{lemma}

\begin{proof}
This lemma follows from the observation that: if $\sigma \to \varrho_1\ldots\varrho_k \in S$, $\varrho \in \mindices$, and $\gamma_2 \in
\mL(\aut_\varrho)$, then replacing any occurrence of $\sigma$ in $\gamma_2$ by $\varrho_1\ldots\varrho_k$ results in $\gamma_2' \in
\mL(\aut_\varrho)$.
\myEnd
\end{proof}

The following lemma is the main result of this subsection.

\begin{lemma} \label{lemma: model graph}
Let $X$ and $\Gamma$ be finite sets of formulas in NCNF of the base language, and let $M = \langle W,R,H\rangle$ be a~consistent and
\cL-saturated model graph such that $\Gamma \subseteq H(w)$, for all $w \in W$, and $X \subseteq H(\tau)$, for some $\tau \in W$. Then the
$L$-model $M'$ corresponding to $M$ validates $\Gamma$ and satisfies $X$ at~$\tau$.
\end{lemma}

\begin{proof}
Observer that:
\begin{align}
\!\!\!\!-&\mbox{ if $[A,q]\psi\! \in\! H(w)$ and
    $R_\sigma(w,w')$ and $\delta_A(q,\sigma,q')$
    then $[A,q']\psi\! \in\! H(w')$}\label{eq:firstshow1}\\
\!\!\!\!-&\mbox{ if $[A,q]\psi\! \in\! H(w)$ and
    $R_{\cnv{\sigma}\!\!}(w'\!,\!w)$ and
    $\delta_A(q,\sigma,q'\!)$ then
    $[A,q']\psi\! \in\! H(w')$.}\label{eq:firstshow2}
\end{align}
These assertions hold because:
\begin{itemize}
\item if $[A,q]\psi \in H(w)$ and $\delta_A(q,\sigma,q')$ holds then $\Box_\sigma[A,q']\psi \in H(w)$
\item if $\Box_\sigma[A,q']\psi \in H(w)$ and $R_\sigma(w,w')$ holds then $[A,q']\psi \in H(w')$
\item if $\Box_\sigma[A,q']\psi \in H(w)$ and $R_{\cnv{\sigma}}(w',w)$ holds then $[A,q']\psi \in H(w')$.
\end{itemize}

Using induction on the construction of $\varphi$, we now prove that for any $u \in W$,\break
if $\varphi \in H(u)$ and $\varphi$ is
a~formula of the base language then $M',u \models \varphi$.
Assume that $\varphi \in H(u)$. The only non-trivial cases are when
\begin{align}
&\mbox{$\varphi = [\alpha]\psi$, or}\label{eq:mainlemma-caseA}\\
&\mbox{$\varphi = \lDmd{\alpha}\psi$, where $\alpha \notin \mindices$ and $\alpha$ is not a~test.}\label{eq:mainlemma-caseB}
\end{align}

Consider the case~\eqref{eq:mainlemma-caseA} and let $q \in I_{\aut_\alpha}$.
By Lemma \ref{lemma: prog-aut}, it suffices to show that $M',u \models [\aut_\alpha,q]\psi$. Since $M$ is \cL-saturated and $\varphi
\in H(u)$, we have that $[\aut_\alpha,q]\psi \in H(u)$. Let $v \in W$ be a~node such that $(u,v) \in (\aut_\alpha,q)^{M'}$. We show that
$\psi \in H(v)$. By the construction of $M'$ and Lemma \ref{lemma: aut-Aut}, there exist a~word $\omega_1\ldots\omega_k \in
\mL((\aut_\alpha,q))$ and elements $w_0, \ldots, w_k$ of $W$ such that $w_0 = u$, $w_k = v$, and for every $1 \leq i \leq k$~:
\begin{align}
&\mbox{$\omega_i \in \mindices$ and $R_{\omega_i}(w_{i-1},w_i)$ or $R_{\cnv{\omega}_i}(w_i,w_{i-1})$ holds, or
}\label{eq:xwg-1}\\
&\mbox{$\omega_i$ is of the form $(\xi_i?)$ and $w_{i-1} = w_i \in \xi_i^{M'}$.
}\label{eq:xwg-2}
\end{align}
Let $q_0 = q$, $q_1$, \ldots, $q_k$ be an accepting run of $(\aut_\alpha,q)$ on the word $\omega_1\ldots\omega_k$. We have that $q_k \in
F_{\aut_\alpha}$. We prove by an inner induction on $0 \leq i \leq k$ that $[\aut_\alpha,q_i]\psi \in H(w_i)$.
The base case $i = 0$ clearly holds.
Inductively, assume that $1 \leq i \leq k$ and $[\aut_\alpha,q_{i-1}]\psi \in H(w_{i-1})$. If \eqref{eq:xwg-1} holds then, by
\eqref{eq:firstshow1} and \eqref{eq:firstshow2}, it follows that $[\aut_\alpha,q_i]\psi \in H(w_i)$. Suppose that \eqref{eq:xwg-2} holds.
Since $[\aut_\alpha,q_{i-1}]\psi \in H(w_{i-1})$ and $M$ is \cL-saturated, we have that $\Box_{(\xi_i?)}[\aut_\alpha,q_i]\psi \in
H(w_{i-1})$. Since $M',w_{i-1} \models \xi_i$, by Lemma \ref{lemma: gen clash}, $\ovl{\xi_i} \notin H(w_{i-1})$. Since $M$ is
\cL-saturated, it follows that $[\aut_\alpha,q_i]\psi \in H(w_{i-1})$. Since $w_{i-1} = w_i$, we also have that $[\aut_\alpha,q_i]\psi \in
H(w_i)$. This completes the inner induction. As a~consequence, $[\aut_\alpha,q_k]\psi \in H(w_k)$. Since $q_k \in F_{\aut_\alpha}$ and $v=
w_k$, it follows that $\psi \in H(v)$. By the inductive assumption, we have that $M',v \models \psi$. Therefore $M',u \models
[\aut_\alpha,q]\psi$.

Consider the case~\eqref{eq:mainlemma-caseB}. Since $M$ is \cL-saturated, there exists $q \in I_{\Aut_\alpha}$ such that
$\lDmd{\Aut_\alpha,q}\psi \in H(u)$.
By Lemma \ref{lemma: prog-aut}, it suffices to show that $M',u \models \lDmd{\Aut_\alpha,q}\psi$. Since $\lDmd{\Aut_\alpha,q}\psi \in
H(u)$, by the global consistency of $M$, there exist an accepting run $q_0 = q$, $q_1$, \ldots, $q_k$ of $(\Aut_\alpha,q)$ on a~word
$\omega_1\ldots\omega_k$ and nodes $w_0 = u$, $w_1$, \ldots, $w_k$ of $M$ such that $\psi \in H(w_k)$ and, for $1 \leq i \leq k$,
$\lDmd{\Aut_\alpha,q_i}\psi \in H(w_i)$ and if $\omega_i \in \mindices$ then $R_{\omega_i}(w_{i-1},w_i)$ holds, else $\omega_i$ is of the
form $(\xi_i?)$ and $w_{i-1} = w_i$ and $\lDmd{\xi_i?}\lDmd{\Aut_\alpha,q_i}\psi \in H(w_{i-1})$. Since $M$ is \cL-saturated, in the case
$\omega_i = (\xi_i?)$ we have that $\xi_i \in H(w_{i-1})$, and by the inductive assumption, it follows that $M',w_{i-1} \models \xi_i$.
Hence $(w_0,w_k) \in (\Aut_\alpha,q)^{M'}$. Since $\psi \in H(w_k)$, by the inductive assumption, we have that $M',w_k \models \psi$.
Hence $M',w_0 \models \lDmd{\Aut_\alpha,q}\psi$. That is, $M',u \models \lDmd{\Aut_\alpha,q}\psi$. This completes the proof.
\myEnd
\end{proof}


\subsection{Completeness}

\begin{Definition}
Let $G$ be a \cL-tableau for $(X,\Gamma)$, $G_m$ be a consistent marking of a $G$, and $v$ be a node of $G_m$. A {\em saturation path} of $v$ w.r.t.\ $G_m$ is a finite sequence $v_0 = v$, $v_1$, \ldots, $v_k$ of nodes of $G_m$, with $k \geq 0$, such that, for every $0 \leq i < k$, $v_i$ is a non-state and $(v_i,v_{i+1})$ is an edge of $G_m$, and $v_k$ is a state.
\end{Definition}

\begin{lemma} \label{lemma: existence of s-p}
Let $G$ be a \cL-tableau for $(X,\Gamma)$ with a consistent marking $G_m$.
Then each node $v$ of $G_m$ has a saturation path w.r.t.~$G_m$. 
\end{lemma}

\begin{proof}
We construct a saturation path $v_0, v_1, \ldots$ of $v$ w.r.t.\ $G_m$ as follows. Set $v_0 := v$ and $i := 0$. While $v_i$ is not a state do: 
\begin{itemize}
\item If the principal of the static rule expanding $v_i$ is not of the form $\lDmd{A,q}\varphi$ then let $v_{i+1}$ be any successor of $v_i$ that belongs to $G_m$ and set $i := i+1$. 
\item If the principal of the static rule expanding $v_i$ is of the form $\lDmd{A,q}\varphi$ then:
  \begin{itemize}
  \item let $v_{i+1},\ldots,v_j$ be the longest sequence of non-states of $G_m$ such that there exist formulas $\varphi_{i+1}$, \ldots, $\varphi_j$ such that the sequence $(v_i,\varphi_i)$, \ldots, $(v_j,\varphi_j)$ is a prefix of a (fixed) $\Dmd$-realization in $G_m$ for $\lDmd{A,q}\varphi$ at $v_i$; 
  \item set $i := j$.
  \end{itemize}
\end{itemize}

The loop terminates because:
\begin{itemize}
\item each formula not of the forms $\lDmd{A,q}\varphi$ and $\lDmd{\psi?}\varphi$ may be reduced at most once
\item each formula of the form $\lDmd{\psi?}\varphi$ is reduced to $\psi$ and $\varphi$ 
\item each formula of the form $\lDmd{A,q}\varphi$ is reduced according to some $\Dmd$-realization. 
\myEnd
\end{itemize}
\end{proof}

We are now in position to prove completeness of the calculus.

\begin{lemma}[Completeness] \label{lemma: comp CPDLreg}
Let $G$ be a \cL-tableau for $(X,\Gamma)$. 
Suppose that $G$ has a~consistent marking $G_m$.
Then $X$ is $L$-satisfiable w.r.t.\ the set $\Gamma$ of global assumptions.
\end{lemma}

\begin{proof}
We construct a~model graph $M = \langle W,R,H\rangle$ as follows:

Let $v_0$ be the root of $G_m$ and $v_0,\ldots,v_k$ be a~saturation path of $v_0$ w.r.t.\ $G_m$. Set $R_\sigma := \emptyset$ for
    all $\sigma \in \mindices$ and set $W := \{\tau\}$, where $\tau$ is a~new node. Set further $H(\tau) := \Label(v_k) \cup \RFormulas(v_k)$.
    Mark $\tau$ as {\em unresolved} and let $f(\tau) = v_k$. (Each node of $M$ will be marked either as {\em unresolved} or as {\em
    resolved}, and $f$ will map each node of $M$ to a state of $G_m$.)

While $W$ contains unresolved nodes, take one unresolved node $w_0$ and do:

 \begin{enumerate}
 \item \label{Step Dmd-realization} For every formula $\lDmd{\sigma}\varphi \in H(w_0)$ (with $\sigma \in \mindices$) do:
 \begin{enumerate}
 \item Let $\varphi_0 = \lDmd{\sigma}\varphi$ and $\varphi_1 = \varphi$.
 \item Let $u_0 = f(w_0)$ and let $u_1$ be the successor of $u_0$ such that $\CELabel(u_1) = \varphi_0$. (As a maintained property of $f$, $u_0$ is a state, $\varphi_0 \in \Label(u_0)$, and therefore $\AfterTrans(u_1)$ holds and $\varphi_1 \in \Label(u_1)$.)

 \item If $\varphi$ is of the form $\lDmd{A,q}\psi$ then let $(u_1,\varphi_1),\ldots,(u_l,\varphi_l)$ to be a~$\Dmd$-realization in $G_m$ for $\varphi_1$ at $u_1$ and let $u_l,\ldots,u_m$ to be a~saturation path of $u_l$ w.r.t.~$G_m$.

 \item Else let $u_1,\ldots,u_m$ be a~saturation path of $u_1$ w.r.t.~$G_m$.

 \item Let $j_0 = 0 < j_1 < \ldots < j_{n-1} < j_n = m$ be all the indices such that, for $0 \leq j \leq m$, $u_j$ is a state of $G$ iff $j \in \{j_0,\ldots,j_n\}$. For $0 \leq s \leq n-1$, let $\lDmd{\sigma_s}\varphi_{j_s+1} = \CELabel(u_{j_s + 1})$. (We have that $\sigma_0 = \sigma$.)

 \item \label{Step create new node} For $1 \leq s \leq n$ do:
    \begin{enumerate}
    \item Let $Z_s = \bigcup \{ \Label(u_i) \cup \RFormulas(u_i) \mid j_{s-1}+1 \leq i \leq j_s \}$.
    \item If there does not exist $w_s \in W$ such that $H(w_s) = Z_s$ then: add a~new node $w_s$ to $W$, set $H(w_s) := Z_s$, mark $w_s$ as unresolved, and let \mbox{$f(w_s) = u_{j_s}$}.
    \item Add the pair $(w_{s-1},w_s)$ to $R_{\sigma_{s-1}}$.
    \end{enumerate}

  \end{enumerate}

 \item Mark $w_0$ as resolved.
 \end{enumerate}

As $H$ is a~one-to-one function and $H(w)$ of each $w \in W$ is a subset of $\clsL(X \cup \Gamma)$, the above construction terminates and results in a~finite model graph.

Observe that, in the above construction we transform the chain $u_0,\ldots,u_m$ of nodes of $G_m$ to a~chain $w_0,\ldots,w_n$ of nodes of $M$ by sticking together nodes in every saturation path. Hence, $M$ is \cL-saturated and satisfies the local consistency property.

Suppose that $\lDmd{A',q'}\psi' \in H(w_0)$. Since $u_0$ is a state of the consistent marking $G_m$, the formula $\lDmd{A',q'}\psi'$ of $H(w_0)$ must have
a~(finite) trace starting from $w_0$, using only $w_0$ (as the first component of the pairs), and ending with a~pair of the form
$(w_0,\psi')$ or $(w_0,\lDmd{\sigma'}\lDmd{A',q''}\psi')$. This together with Step~\ref{Step Dmd-realization} implies that $M$ satisfies
the global consistency property. Hence, $M$ is a~consistent and \cL-saturated model graph.

Since $\Label(v_0) = X \cup \Gamma$ and $\Label(v_0) \subseteq \Label(v_k) \cup \RFormulas(v_k)$, we have that $X \subseteq H(\tau)$ and $\Gamma
\subseteq H(\tau)$. Consider Step~\ref{Step create new node} of the construction, as $u_{j_{s-1}}$ is an ``and''-node and $u_{j_{s-1} + 1}$
is a~successor of $u_{j_{s-1}}$ that is created by the transitional rule, we have that $\Label(u_{j_{s-1} + 1}) \supseteq \Gamma$, and
hence $\Label(u_{j_s}) \cup \RFormulas(u_{j_s}) \supseteq \Label(u_{j_{s-1} + 1}) \supseteq \Gamma$. Hence $\Gamma \subseteq H(w_s)$ for every
$w_s \in W$. By Lemma \ref{lemma: model graph}, the Kripke model corresponding to $M$ validates $\Gamma$ and satisfies $X$ at $\tau$.
Hence, $X$ is $L$-satisfiable w.r.t.~$\Gamma$.
\myEnd
\end{proof}
} 


\section{An ExpTime Tableau Decision Procedure for \CPDLreg}
\label{section: proc}

Let $G = (V,E,\nu)$ be a \cL-tableau for $(X,\Gamma)$ and $G_m$ be a marking of $G$.
%
The {\em graph $G_t$ of traces of $G_m$ in $G$} is defined as follows:
\begin{itemize}
\item nodes of $G_t$ are pairs $(v,\varphi)$, where $v \in V$ and $\varphi \in \Label(v)$
\item a~pair $((v,\varphi),(w,\psi))$ is an edge of $G_t$
    if $v$ is a~node of $G_m$, $\varphi$ is of the form
    $\lDmd{A,q}\xi$ or
    $\lDmd{\omega}\lDmd{A,q}\xi$, and the
    sequence $(v,\varphi)$, $(w,\psi)$ is a~subsequence of a
    trace in $G_m$.
\end{itemize}
 A~node $(v,\varphi)$ of $G_t$ is an {\em end node} if $\varphi$ is a~formula of the base
 language. A~node of $G_t$ is {\em productive} if there is a~path connecting it to an end node.

\begin{algorithm}[t!]
\caption{for checking $L$-satisfiability of $X$ w.r.t.~$\Gamma$.\label{alg1}}
\Input{finite sets $X$ and $\Gamma$ of formulas in NCNF of the base language, the mapping $\aut$ specifying the finite automata of the symmetric 
regular semi-Thue system of the considered \CPDLreg logic $L$}
\Output{{\em true} if $X$ is $L$-satisfiable w.r.t.\ $\Gamma$, and {\em false} otherwise}

\BlankLine
let $G = (V,E,\nu)$ be the \cL-tableau constructed by $\Tableau(X,\Gamma)$ (using $\aut$)\;

\BlankLine
\While{$\Status(\nu) \neq \Unsat$}
{
  let $G_m$ be the subgraph of $G$ induced by the nodes with status different from $\Unsat$ and $\Incomplete$\tcc*[l]{we have that $G_m$ is a~marking of $G$}

  \BlankLine
  construct the graph $G_t$ of traces of $G_m$ in $G$\;

  \BlankLine
  \uIf{there exist a node $v$ of $G_m$ and a formula $\lDmd{A,q}\varphi \in \Label(v)$ such that $(v,\lDmd{A,q}\varphi)$ is not a productive node of~$G_t$}{$\Status(v) := \Unsat$,\ \ $\PropagateStatus(v)$}\label{alg1-step-comp-nodes-not-satisfying-global-consistency}
  \lElse{\Return $\True$}
}
\Return $\False$\label{alg1:step-last}\;
\end{algorithm}

Algorithm~\ref{alg1} (given on page~\pageref{alg1}) is a simple \EXPTIME algorithm for checking $L$-satisfiability of $X$ w.r.t.~$\Gamma$. \ShortVersion{See~\cite{nCPDLreg-long} for the proof of the following theorem.}

\LongVersion{
\begin{lemma} \label{lemma: alg1 compl}
Let $h = |\Sigma|$ and let 
$k$ be the sum of the sizes of the automata $\aut_\sigma$ (for $\sigma \in \Sigma$), and $l$ be the size of $X \cup \Gamma$. 
Then Algorithm~\ref{alg1} for $X$, $\Gamma$, $\aut$ runs in $2^{O(n)}$ steps, where $n$ be the size of $\clsL(X \cup \Gamma)$ and is polynomial in $k.h.l$.
\end{lemma}

\begin{proof}
By Lemma~\ref{lemma: properties of tableaux}, $n$ is polynomial in $k.h.l$ and the graph $G$ has $2^{O(n)}$ nodes and can be constructed in $2^{O(n)}$.
For each node $v$ of $G$, $\Label(v) \subseteq \clsL(X \cup \Gamma)$. 
Constructing graphs of traces and updating statuses of nodes can be done in polynomial time in the size of $G$. Hence, totally, Algorithm~\ref{alg1} runs in $2^{O(n)}$ steps.
\myEnd
\end{proof}
} 

\begin{theorem} \label{theorem: FFGSD}
Let $S$ be a~symmetric regular semi-Thue system over $\Sigma$,
$\aut$ be the mapping specifying the finite automata of $S$,
and $L$ be the \CPDLreg logic corresponding to $S$. Let $X$ and
$\Gamma$ be finite sets of formulas in NCNF of the base
language. Then Algorithm~\ref{alg1} is an \EXPTIME (optimal) decision
procedure for checking whether $X$ is $L$-satisfiable w.r.t.\
the set $\Gamma$ of global assumptions.
\end{theorem}

\LongVersion{
\begin{proof}
It is straightforward to prove by induction that the algorithm has the invariant that a~consistent marking of $G$ cannot contain any node with status $\Unsat$. 
For the base step, use Theorem~\ref{theorem: s-c} and the local consistency property. For the induction step, use Theorem~\ref{theorem: s-c} and the global consistency property. 

The algorithm returns $\False$ only when $\Status(\nu) = \Unsat$, i.e., only when $G$ does not have any consistent marking. At Step~7, $G_m$ is a consistent marking of $G$. That is, the algorithm returns $\True$ only when $G$ has a consistent marking. Therefore, by Theorem~\ref{theorem: s-c}, Algorithm~\ref{alg1} is a~decision procedure for the considered problem. The complexity was established by Lemma~\ref{lemma: alg1 compl}.
\myEnd
\end{proof}
}

Clearly, one may modify procedure $\Tableau$ to make it stop as soon as $\Status(\nu) = \Unsat$. We excluded this condition just to formalize and prove Theorem~\ref{theorem: s-c}, but it does not affect correctness of Theorem~\ref{theorem: FFGSD}.


\section{Checking Global Consistency On-the-Fly}
\label{section: cgc otf}

Observe that Algorithm~\ref{alg1} first constructs a \cL-tableau and then checks whether the tableau contains a~consistent marking. To speed up the performance these two tasks can be done concurrently. For this we can update $G_m$ and $G_t$ during the construction of $G$ and detect and propagate $\Unsat$ (and $\Sat$) on-the-fly using both the local and global consistency properties as discussed below.

For nodes $(v,\varphi)$ of the graph of traces $G_t$, define $\SemiEndNodes(v,\varphi)$ to be the smallest sets of nodes of $G_t$ that satisfy the following conditions:
\begin{itemize}
\item if $(v,\varphi)$ is an end node of $G_t$ or $\Status(v) \in \{\Unexpanded, \Sat\}$ then\\ $\SemiEndNodes(v,\varphi) = \{(v,\varphi)\}$, 
\item else $\SemiEndNodes(v,\varphi) = \bigcup\,\{\SemiEndNodes(w,\psi) \mid (w,\psi)$ is a successor of $(v,\varphi)$ in $G_t$ different from $(v,\varphi)\} \setminus \{(v,\varphi)\}$.
\end{itemize}

Observe that, during the construction of $G$ and $G_t$, if $\SemiEndNodes(v,\varphi) = \emptyset$ then $(v,\varphi)$ is not and will never be a productive node of $G_t$ and hence $\Status(v)$ can be set to $\Unsat$. Furthermore, at the end (i.e., when $G$ is a full ``and-or'' graph, $G_m$ is the subgraph of $G$ induced by the nodes with status different from $\Unsat$ and $\Incomplete$, and $G_t$ is the graph of traces of $G_m$ in $G$), if a node $(v,\varphi)$ of $G_t$ has $\SemiEndNodes(v,\varphi) \neq \emptyset$ then $(v,\varphi)$ is a productive node of $G_t$.

An improved \EXPTIME tableau decision procedure for checking $L$-satisfiability of $X$ w.r.t.~$\Gamma$ is as follows. During the construction of a \cL-tableau for $(X,\Gamma)$:
\begin{itemize}
\item let $G = (V,E,\nu)$ be the current (partial) ``and-or'' graph; 
\item update $G_m$ on-the-fly so that it is the subgraph of $G$ induced by the nodes with status different from $\Unsat$ and $\Incomplete$;
\item update $G_t$ on-the-fly so that it is the graph of traces of $G_m$ in $G$;
\item update $\SemiEndNodes(v,\varphi)$ on-the-fly for nodes $(v,\varphi)$ of $G_t$ as follows:
  \begin{itemize}
  \item when a new node $v$ with status $\Unexpanded$ is created for $G$, for every node $(v,\varphi)$ created for $G_t$, set $\SemiEndNodes(v,\varphi) := \{(v,\varphi)\}$,
  \item when a node $v$ of $G$ receives status $\Sat$, for every node $(v,\varphi)$ of $G_t$, set $\SemiEndNodes(v,\varphi) := \{(v,\varphi)\}$,
  \item when a node $v$ of $G$ is expanded and receives status $\Expanded$, 
	\begin{itemize}
	\item for every end node $(v,\varphi)$ of $G_t$, set $\SemiEndNodes(v,\varphi) := \{(v,\varphi)\}$,
	\item for every non-end node $(v,\varphi)$ of $G_t$, set 
	\begin{equation}\label{eq: JHDSA}
	\parbox{12cm}{$\SemiEndNodes(v,\varphi) := \bigcup\,\{\SemiEndNodes(w,\psi) \mid (w,\psi)$ is a~successor of $(v,\varphi)$ in $G_t$ different from $(v,\varphi)\} \setminus \{(v,\varphi)\}$;}
	\end{equation}
	\end{itemize}
  \item when a node $(w,\psi)$ of $G_t$ has $\SemiEndNodes(w,\psi)$ changed, for every predecessor $(v,\varphi)$ of $(w,\psi)$ in $G_t$, update $\SemiEndNodes(v,\varphi)$ by~\eqref{eq: JHDSA}; 
  \item when a node $w$ of $G$ receives status $\Unsat$ or $\Incomplete$ or is deleted, for every node $(v,\varphi)$ of $G_t$ that is a predecessor of some $(w,\psi)$, update $\SemiEndNodes(v,\varphi)$ by~\eqref{eq: JHDSA};   
  \end{itemize}
\item when a node $(v,\varphi)$ of $G_t$ receives $\SemiEndNodes(v,\varphi) = \emptyset$, set $\Status(v) := \Unsat$ and execute $\PropagateStatus(v)$;
\item when the root $\nu$ of $G$ receives status $\Unsat$, return $\False$ (which means $X$ is not $L$-satisfiable w.r.t.~$\Gamma$);
\item at the end (not terminated as in the above case), return $\True$ (which means $X$ is $L$-satisfiable w.r.t.~$\Gamma$).
\end{itemize}


See also~\cite{Nguyen08CSP-FI} for other possible optimization techniques, which have been implemented and evaluated for the description logic \ALC. 



\section{Conclusions}
\label{section: conc}

As discussed in the introduction, translating the general satisfiability checking problem in \CPDLreg into the satisfiability checking problem in \CPDL may increase the complexity (and may decrease efficiency). Therefore, the direct approach for automated reasoning in \CPDLreg is worth studying. 

In this paper we have given a tableau calculus and the first cut-free \EXPTIME tableau decision procedure for checking satisfiability in the logic \CPDLreg. Our decision procedure uses global caching for states and local caching for non-states in tableaux. In comparison with the tableau calculus given in our joint papers~\cite{GBGI,dkns2011} for \CPDLreg, the tableau calculus given in the current paper is essentially better since it is cut-free. Furthermore, we have explicitly incorporated on-the-fly propagation of local (in)consistency into the latter calculus and the corresponding decision procedure. We have also discussed how on-the-fly propagation of global (in)consistency can be incorporated into the tableau decision procedure. 

When restricted to \CReg, our tableau decision procedure is much better than the one proposed by us and Sza\l{}as in~\cite{NguyenSzalas-CADE-22} for \CReg, as it does not use cuts and it does not have to check the global consistency property (this is an effect of cut elimination).

On technical matters, apart from global caching of states as in the works~\cite{GoreW09,GoreW10} by Gor{\'e} and Widmann, we allow to cache also non-states. We apply global caching for nodes in the local graphs of non-states $v$ satisfying $\AfterTrans(v)$. Such local graphs in~\cite{GoreW09,GoreW10} are trees. We propose not to delay solving incompatibility w.r.t. converse as long as in \cite{GoreW09,GoreW10}.\footnote{Recall the example with $\{p, \lDmd{\sigma}[\sigma^-]\lnot p, [\sigma]((p_1 \lor q_1) \land\ldots\land(p_n \lor q_n))\}$.} One can solve incompatibility w.r.t. converse as soon as possible as done in this paper. By giving a higher priority to the current branch even when it involves converse we can further favor depth-first search. 


\begin{thebibliography}{10}

\bibitem{BaaderSattler01}
F.~Baader and U.~Sattler.
\newblock An overview of tableau algorithms for description logics.
\newblock {\em Studia Logica}, 69:5--40, 2001.

\bibitem{de-giacomo-massacci-converse-pdl}
G.~{De Giacomo} and F.~Massacci.
\newblock Combining deduction and model checking into tableaux and algorithms
  for {Converse-PDL}.
\newblock {\em Information and Computation}, 162(1-2):117--137, 2000.

\bibitem{Demri01}
S.~Demri.
\newblock The complexity of regularity in grammar logics and related modal
  logics.
\newblock {\em Journal of Logic and Computation}, 11(6):933--960, 2001.

\bibitem{DemriNivelle2004}
S.~Demri and H.~de~Nivelle.
\newblock Deciding regular grammar logics with converse through first-order
  logic.
\newblock arXiv:cs.LO/0306117, 2004.

\bibitem{ddn-jlli05}
S.~Demri and H.~de~Nivelle.
\newblock Deciding regular grammar logics with converse through first-order
  logic.
\newblock {\em Journal of Logic, Language and Inform.}, 14(3):289--329, 2005.

\bibitem{GBGI}
B.~Dunin-K\c{e}plicz, L.A. Nguyen, and A.~Sza\l{}as.
\newblock A framework for graded beliefs, goals and intentions.
\newblock {\em Fundam. Inform.}, 100(1-4):53--76, 2010.

\bibitem{HSPDL}
B.~Dunin-K\c{e}plicz, L.A. Nguyen, and A.~Sza{\l}as.
\newblock Tractable approximate knowledge fusion using the {Horn} fragment of
  serial propositional dynamic logic.
\newblock {\em Int. J. Approx. Reasoning}, 51(3):346--362, 2010.

\bibitem{dkns2011}
B.~Dunin-K\c{e}plicz, L.A. Nguyen, and A.~Sza{\l}as.
\newblock {Converse-PDL} with regular inclusion axioms: a framework for {MAS}
  logics.
\newblock {\em Journal of Applied Non-Classical Logics}, 21(1):61--91, 2011.

\bibitem{GiacomoL96}
G.~De Giacomo and M.~Lenzerini.
\newblock {TBox} and {ABox} reasoning in expressive description logics.
\newblock In L.C. Aiello, J.~Doyle, and S.C. Shapiro, editors, {\em Proceedings
  of KR'1996}, pages 316--327. Morgan Kaufmann, 1996.

\bibitem{Gore99}
R.~Gor\'{e}.
\newblock Tableau methods for modal and temporal logics.
\newblock In D'Agostino et~al, editor, {\em Handbook of Tableau Methods}, pages
  297--396. Kluwer, 1999.

\bibitem{GoreNguyen05tab}
R.~Gor{\'e} and L.A. Nguyen.
\newblock A tableau system with automaton-labelled formulae for regular grammar
  logics.
\newblock In B.~Beckert, editor, {\em Proceedings of TABLEAUX 2005, LNAI 3702},
  pages 138--152. Springer-Verlag, 2005.

\bibitem{GoreNguyenCLIMA07}
R.~Gor{\'e} and L.A. Nguyen.
\newblock Analytic cut-free tableaux for regular modal logics of agent beliefs.
\newblock In F.~Sadri and K.~Satoh, editors, {\em Proceedings of CLIMA VIII},
  pages 274--289, 2007.

\bibitem{GoreNguyenTab07}
R.~Gor\'{e} and L.A. Nguyen.
\newblock {EXPTIME} tableaux with global caching for description logics with
  transitive roles, inverse roles and role hierarchies.
\newblock In N.~Olivetti, editor, {\em Proc.\ of TABLEAUX 2007, LNAI 4548},
  pages 133--148. Springer-Verlag, 2007.

\bibitem{GoreW09}
R.~Gor{\'e} and F.~Widmann.
\newblock Sound global state caching for $\mathcal{ALC}$ with inverse roles.
\newblock In M.~Giese and A.~Waaler, editors, {\em Proceedings of TABLEAUX
  2009}, volume 5607 of {\em LNCS}, pages 205--219. Springer, 2009.

\bibitem{GoreW10}
R.~Gor{\'e} and F.~Widmann.
\newblock Optimal and cut-free tableaux for propositional dynamic logic with
  converse.
\newblock In J.~Giesl and R.~H{\"a}hnle, editors, {\em Proceedings of IJCAR
  2010}, volume 6173 of {\em LNCS}, pages 225--239. Springer, 2010.

\bibitem{HKT00}
D.~Harel, D.~Kozen, and J.~Tiuryn.
\newblock {\em Dynamic Logic}.
\newblock MIT Press, 2000.

\bibitem{HorrocksKS06}
I.~Horrocks, O.~Kutz, and U.~Sattler.
\newblock The even more irresistible $\mathcal{SROIQ}$.
\newblock In P.~Doherty, J.~Mylopoulos, and C.A. Welty, editors, {\em Proc.\ of
  KR'2006}, pages 57--67. AAAI Press, 2006.

\bibitem{MS97}
A.~Mateescu and A.~Salomaa.
\newblock Formal languages: an introduction and a synopsis.
\newblock In {\em Handbook of Formal Languages - Volume~1}, pages 1--40.
  Springer, 1997.

\bibitem{nguyen01B5SL}
L.A. Nguyen.
\newblock Analytic tableau systems and interpolation for the modal logics {KB},
  {KDB}, {K5}, {KD5}.
\newblock {\em Studia Logica}, 69(1):41--57, 2001.

\bibitem{Nguyen08CSP-FI}
L.A. Nguyen.
\newblock An efficient tableau prover using global caching for the description
  logic $\mathcal{ALC}$.
\newblock {\em Fundamenta Informaticae}, 93(1-3):273--288, 2009.

\bibitem{NguyenSzalas-KSE09}
L.A. Nguyen and A.~Sza\l{}as.
\newblock An optimal tableau decision procedure for {Converse-PDL}.
\newblock In N.-T. Nguyen, T.-D. Bui, E.~Szczerbicki, and N.-B. Nguyen,
  editors, {\em Proceedings of KSE'2009}, pages 207--214. IEEE Computer
  Society, 2009.

\bibitem{NguyenSzalas-CADE-22}
L.A. Nguyen and A.~Sza\l{}as.
\newblock A tableau calculus for regular grammar logics with converse.
\newblock In Renate~A. Schmidt, editor, {\em Proceedings of CADE-22}, volume
  5663 of {\em LNAI}, pages 421--436. Springer-Verlag, 2009.

\bibitem{NguyenS10FI}
L.A. Nguyen and A.~Sza\l{}as.
\newblock Checking consistency of an {ABox} w.r.t. global assumptions in {PDL}.
\newblock {\em Fundamenta Informaticae}, 102(1):97--113, 2010.

\bibitem{Pratt80}
V.R. Pratt.
\newblock A near-optimal method for reasoning about action.
\newblock {\em J.~Comp.~Syst.~Sci.}, 20(2):231--254, 1980.

\bibitem{Rautenberg83}
W.~Rautenberg.
\newblock Modal tableau calculi and interpolation.
\newblock {\em JPL}, 12:403--423, 1983.

\bibitem{Vardi98}
M.Y. Vardi.
\newblock Reasoning about the past with two-way automata.
\newblock In K.G. Larsen, S.~Skyum, and G.~Winskel, editors, {\em Proceedings
  of ICALP'98}, volume 1443 of {\em LNCS}, pages 628--641. Springer, 1998.

\end{thebibliography}
\end{document}